\documentclass[a4paper,11pt]{article}

\usepackage{jinstpub} 
\usepackage{verbatim}
\usepackage[compact]{titlesec}
\usepackage{enumitem}
\usepackage{multirow}

\usepackage{xcolor}

\usepackage[version=4]{mhchem}

\usepackage{lineno}

\title{\boldmath Gas contamination and mitigation in a 100~m$^3$ / 10~bar argon TPC with optical readout: a viability study}

\author[a,b]{D. J. Fernández-Posada,}
\author[a,*]{D. González-Díaz,}
\author[b]{J. Baldonedo,}
\author[b]{J. Collazo,}
\author[b]{E. Casarejos,}
\author[c]{P. Hamacher-Baumann,}
\author[a]{P. Amedo,}
\author[a]{J. Llerena}

\affiliation[a]{Instituto Galego de Física de Altas Energías, Univ. de Santiago de Compostela, Campus sur, Rúa Xosé María Suárez Núñez, s/n, Santiago de Compostela, 15782, Spain}
\affiliation[b]{CINTECX, Universidade de Vigo, Dpt. Mech. Engineering, 36310 Vigo, Spain}
\affiliation[c]{III. Physikalisches Institut, RWTH Aachen University, 52056 Aachen, Germany}

\emailAdd{diego.gonzalez.diaz@usc.es}

\abstract{Gaseous Optical Time Projection Chambers (TPCs) aimed at Neutrino Physics and Rare Event Searches will likely exceed the tonne scale during the next decade. This will make their performance sensitive to gas contamination levels as low as 100~ppb, { which} is challenging at room temperature due to outgassing from structural materials. In this work we discuss gas distribution and impurity mitigation { in} a 5~m-length/5~m-diameter 10~bar TPC filled with Ar/CF$_4$ admixed at 99/1 per volume (1.75~tonne), loaded with technical plastics in order to enhance light collection and readout. Different distributor topologies, outgassing and flow rates are discussed.
Specifically, our work is aimed at illustrating the conceptual viability of the optical readout of ND-GAr's { (Near Detector - Gaseous Argon)}
TPC (within the DUNE Near-Detector complex), in terms of material compatibility. 
For our proposal, with perforated distributors aligned with the electric field, and under realistic assumptions, the concentration of contaminants can be controlled within a week { after} chamber filling. 
In the case of N$_2$, injection of fresh gas at \%-level seems to represent the safest strategy to keep the concentration within operability limits.
}


\arxivnumber{2501.05791} 

\begin{document}
\maketitle
\flushbottom

\section{Introduction}

Gas distribution and purification, together with material selection, are crucial aspects for the operation of gaseous detectors (e.g., \cite{Fabio}). 
{%
With future large-volume detectors (in particular time projection chambers, or TPCs) poised to reach the tonne scale in gas mass for the first time, the associated challenges will increase \cite{Dave, Hilke, DGD}. }
{ DUNE, NEXT and CYGNUS-1000 are several examples where the use of very large gaseous TPCs is foreseen.}
The DUNE experiment (dedicated to the study of neutrino oscillations) is considering the construction of a nearly 2~tonne argon-rich TPC.
This TPC would operate with { gaseous argon (GAr)} at 10~bar, with 5~m diameter and 5~m drift as part of the { Near Detector (ND) complex -- such a detector system is known as ND-GAr} \cite{NDGAr, NDGAr-PhaseII}.
The NEXT experiment (targeting the discovery of $\beta\beta0\nu$) aims at a bit over 1~tonne of $^{136}$Xe in a 15~bar TPC of 2.6~m (diameter)$ ~\times 2.6$~m (drift) \cite{NEXT}.
CYGNUS-1000 (whose goal is the reconstruction of the direction of galactic WIMP dark matter) plans operation under He/SF$_6$(CF$_4$) at around atmospheric pressure in a 1000~m$^3$ detector, totaling 3~tonne \cite{CYGNUS}. The need for such unprecedented masses of gas is driven by the unique capability of this phase to resolve the details of the interactions involving low-energy particle tracks, provided a sufficient interaction/decay probability can be guaranteed in the first place. { Time Projection Chambers are based on the principle of drifting ionization electrons over distances of several meters onto a suitably pixelated anode: the detector obtains, additionally, the longitudinal position of these electrons from their arrival times, upon multiplication by the drift velocity $v_d$ (a property of the gas at a given electric field). Minute concentrations of gas pollutants can trap the electrons during their drift and, when the device is used for precise mm-level tracking, they can also cause modifications of the drift velocity at a mere $10^{-4}$ level, significantly distorting the reconstructed particle trajectories.}

{ Gas distribution becomes complicated at the aforementioned scale, from practical considerations alone}: some of the gases considered (e.g., SF$_6$, CF$_4$) represent a threat to the environment,\footnote{Available estimates of the global warming potential (GWP) exceed 20000 for SF$_6$ and 5000 for CF$_4$ \cite{GWP}.} while others (e.g., $^{136}$Xe) are too precious to be vented. { In fact, due to the sheer amount of gas mass, it can be argued that virtually any open loop scheme will become impractical and uneconomical for tonne-scale TPCs, compared to a closed-loop implementation}. Closed-loop operation binds the system performance to the power of the compressors and purification systems, in competition with the outgassing rate from internal elements. When high gas mass is achieved through pressurization, the attachment probability (electron lifetime) is increased (decreased) as $\sim{P}^2$ \cite{Huk} and both charge collection and mm-level space resolutions can be compromised even in the presence of tiny concentrations of contaminants. As shown in this work, the sensitivity of the drift velocity to impurities, combined with experimental values of the attachment coefficient, sets the tolerable H$_2$O and O$_2$ levels at the 0.1-1 parts per million (ppm) scale %
for $\mathcal{O}(10~$m) drift. This is at least one order of magnitude below the purity requirements of the ALICE TPC, the largest gaseous TPC to date \cite{ALICE_TPC}.


Simulations based on computational fluid dynamics (CFD) are nowadays possible in a variety of systems. As they provide critical information on the flow characteristics and the admixing of species, they can reduce the investment in physical prototypes and help anticipate problems. CFD simulations of gaseous detectors usually focus on characteristics of the gas flow such as gas stagnancy (`pockets') and the impact of temperature gradients, since they modify the detector response locally \cite{REF2_ALICE}. However, not much has been done to address the role of contaminants, to the best of the authors' knowledge. Modeling the dynamics of contaminants in fluid detectors is not alien to modern technologies, and it has been successfully done for instance for liquid argon TPCs based on a set of differential equations describing each of the elements in the system \cite{LArCont}.

In this work we study gas distribution and impurity mitigation in the context of future large volume/large mass TPCs intended for experiments in the field of Neutrino Physics and Rare Event Searches. 
We discuss, in particular, a scenario where a massive use of technical plastics would be made in order to enhance the optical response of a TPC aiming at the study of neutrino interactions, following the conceptual proposal outlined in \cite{papersOnArCF4}. 

The structure of this document is as follows. First, section \ref{InputsSim} discusses the main inputs used in the CFD simulations: (\ref{ImpuritiesInTPC}) the influence of contaminants on the TPC response, obtained from existing data or electron-transport simulations; (\ref{outgas}) outgassing rates of technical plastics; and (\ref{GasScheme}) the gas distribution scheme. Section \ref{CFD} deals with the optimization of the geometry of the distributor rods (\ref{sectionRods}), and the introduction of the CFD simulation framework (\ref{CFD_simuls}). Section \ref{results} presents our main results, chiefly the spatial and temporal distributions of common outgassing species in the TPC for different distribution schemes, species and conditions. Finally, we end with our conclusions in section \ref{conclusion}.

\section{Inputs} \label{InputsSim} 

\begin{figure}[h!!!]
       \centering
        \includegraphics[width=1\linewidth]{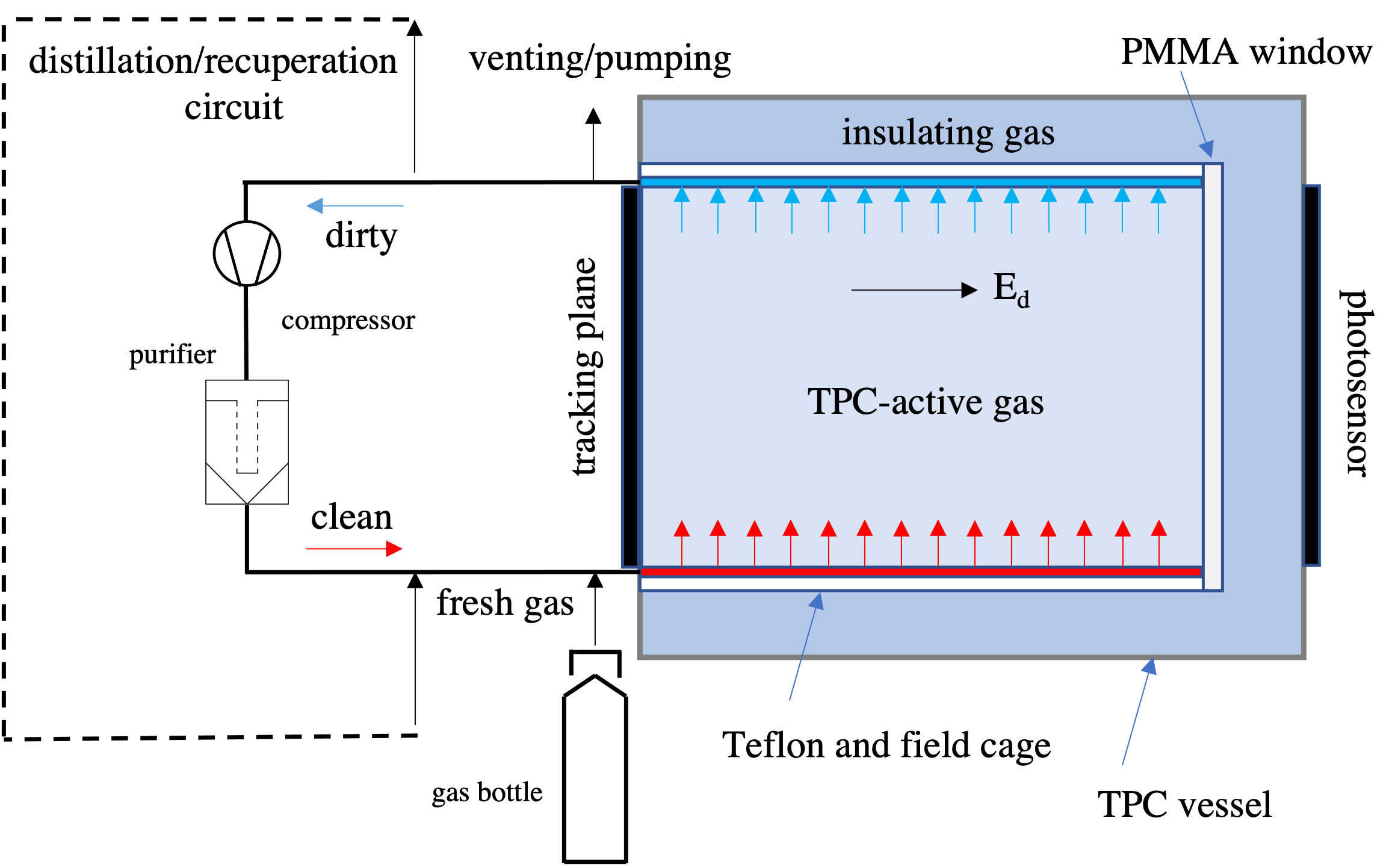}
    \caption{A minimalistic concept of the gas distribution proposed for DUNE's ND-GAr, containing the main elements discussed in text. Manual and automated valves, probing points to assess the gas concentration (and possibly purity), as well as pressure and vacuum gauges will be needed in the final implementation. An additional insulating layer based on gas has been foreseen, acting as electrical (for the field cage) and thermal (for the photosensor) barriers. The full concept will be discussed elsewhere.}
    \label{Gsystem}
\end{figure}

We adopt as a reference geometry a 5 m-length/5 m-diameter cylindrical TPC, similar to the one under consideration for DUNE's ND-GAr \cite{NDGAr, NDGAr-PhaseII}, intermediate in size between \cite{NEXT} and \cite{CYGNUS}. The recently introduced possibility of performing `Full3D' optical tracking in Ar/CF$_4$ (99/1) \cite{papersOnArCF4, CF4Yields, Amedo_spe, Hafeji_tra}, involves the use of large amounts of reflective (Teflon/PTFE) and transparent (PMMA) plastics, to enhance the optical response of the detector. { Teflon, in particular, can intrinsically achieve a reflectivity well above 95\% in the visible region of the spectra, allowing a flatter light collection in the range 35\%~(anode)-65\%~(cathode) and up to $\times 7$ increased light collection for events coming from the distant-most (anode) region \cite{papersOnArCF4}. A diagram of the system and a conceptual gas distribution scheme are shown in Fig. \ref{Gsystem}.}

{
It is assumed that the gas is continuously recirculated through compressors which, along with purifiers (getters), maintain the injected gas clean. Commercial `cold' getters are capable of maintaining O$_2$ and H$_2$O concentrations at ppb-level per pass, while ignoring common molecular additives such as CH$_4$ and CF$_4$
(a getter fulfilling these requirements is model 905 in \cite{SAES1}, for instance). In simulation, since purification levels are at least two orders of magnitude below harmful concentrations, we assume the injected gas is always 100\% pure. Conversely, getters that trap N$_2$ require heating, which may be incompatible with high gas flows (due to accelerated cooling of the active materials of the getter) and molecular additives (due to higher reactivity compared to noble gases). For the nitrogen contamination studies performed here, we neglect the possibility of purification.\footnote{An interesting discussion on the operation and performance of both types of getters can be found in \cite{Monra_Tech}.}}

\subsection{Impact of impurities on TPC response}\label{ImpuritiesInTPC}


Contaminants such as H$_2$O, O$_2$ and N$_2$ will be considered, as they predominantly originate from outgassing of low-vapor-pressure materials, excluding most epoxies and adhesives from detector construction. We disregard contamination from exogenous impurities, such as residual gas in bottles, which could be harmful at concentrations far below those considered here \cite{SF6Anton}. CO$_2$ and CO contaminants can also originate from technical plastics, with outgassing rates comparable to the previously mentioned species \cite{Thieme}. Their impact on chamber performance can be upper-bounded using the H$_2$O case, as it exhibits the strongest electron-cooling power among the discussed molecules. This results in a larger distortion of the drift of the primary ionization electrons compared to the purified gas. Impurities limit TPC performance mainly through: i) the capture of the drifting electrons (attachment) and ii) the alteration of the drift velocity and the ensuing track distortions, affecting momentum reconstruction. For TPCs that rely on scintillation, the one discussed here, iii) modifications to the scintillation probability and absorption can occur too.
Reference values for the contaminant thresholds as used in this paper can be determined by examining the effects in attachment, drift velocity, and light quenching and absorption.

\subsubsection*{Attachment}
Of utmost importance at high pressure is the attachment probability per unit length for electrons drifting in the presence of oxygen ($\eta$), or its related magnitude: the electron lifetime ($\tau_e$). For argon-based mixtures those may be parameterized, after \cite{Huk}, as:
\begin{eqnarray}
    \eta & = & \frac{1}{v_d} \cdot \left( 1 + \frac{f_{H_2O}}{1000} \right) \cdot P^2 \cdot f_{O_2} \cdot \mathcal{A}_{\text{gas}}, \label{att_Huk1} \\
    \tau_e & = & \frac{1}{\eta \cdot v_d}, \label{att_Huk2}
\end{eqnarray}
where the magnitude $\mathcal{A}_{\text{gas}}$ is gas-mixture dependent, $f_{O_2(H_2O)}$ represents the fraction of O$_2$~(H$_2$O) molecules, $v_d$ is the electron drift velocity and $P$ is the gas pressure. The $P^2$-dependence reflects the fact that the dominant attachment process takes place in two steps (Bloch-Bradbury-Herzengber mechanism, BBH, \cite{BBH1, BBH2}):
\begin{eqnarray}
e^- + \textnormal{O}_2 & \rightarrow & \textnormal{O}_2^{-,*}, \\
\textnormal{O}_2^{-,*} + \textnormal{O}_2 ~ (\textnormal{H}_2\textnormal{O}) & \rightarrow & \textnormal{O}_2^- + \textnormal{O}_2^* ~(\textnormal{H}_2\textnormal{O}^*). \label{BBH_H2O} 
\end{eqnarray}
Both $\mathcal{A}_{\text{gas}}$ and $v_d$ depend solely on the pressure-reduced electric field ($E/P$), therefore not showing separate $E$ and $P$ dependencies.\footnote{This is a direct consequence of the fact that $E/P$ determines uniquely the electron energy distribution.} Typical pressure-reduced drift fields in TPCs are found in the range of 10 -- few 100's~V/cm/bar, which are expected to be below the dissociative-attachment thresholds for either CF$_4$, O$_2$ or H$_2$O. This observation has been backed with the state-of-the-art Pyboltz transport code \cite{Pyboltz}, thus anticipating BBH to be the dominant attachment mechanism in the conditions discussed in this work.

The attachment coefficient, $\eta$, in Ar/CF$_4$ (99/1) at $E/P=100$~V/cm/bar can be estimated starting from the experimental $\mathcal{A}_{\text{gas}}$ value in case of the classical Ar/CH$_4$ mixture at 90/10 (`P10'). The compilation in \cite{Huk}, covering pressures up to 4~bar, reports 
{$\mathcal{A}_{\text{gas}}= (15.1 \pm 1.5)\,{\mu}$s$^{-1}~ 
 \textnormal{bar}^{-2}$}, and $v_d = 5.36$ cm/$\mu${s}. Extrapolating to 10~bar following eq. \ref{att_Huk1}, this value would correspond to 10\% charge loss over 5~m drift for O$_2$ concentrations of 0.7~ppm. The same work reports a linear decrease of $\mathcal{A}_{\text{gas}}$ with the decrease of the concentration of the molecular additive and an increase with the size of the molecule (e.g., from CH$_4$ to i-C$_4$H$_{10}$). The effect can be attributed to the dependence of the stabilization reaction \ref{BBH_H2O} with the concentration of the molecular additive and with the molecule size. For Ar/i-C$_4$H$_{10}$ at 99/1, 
{ $\mathcal{A}_{\text{gas}}$ = ($48.4 \pm 3.7)\,{\mu}$s$^{-1} ~\textnormal{bar}^{-2}$}
and $v_d=4.97$ cm/$\mu${s} were measured at $E/P=100$ V/cm/bar, yielding a 10\% charge drop over 5~m for 0.2~ppm of O$_2$. Given that i-C$_4$H$_{10}$ and CF$_4$ molecules are similar in size, a similar strength of reaction \ref{BBH_H2O}, and thus similar attachment rates, might be expected.

An $E/P$ of 100~V/cm/bar would correspond to an imposing 500~kV cathode bias on a 5~m drift distance at 10~bar. Based on the studies of breakdown voltage for Ar/CF$_4$ (99/1) carried out in \cite{LeslieVbd}, as well as the availability of commercial feedthroughs, we will restrict our discussion to a cathode bias not greater than 250~kV ($E/P=50$~V/cm/bar). That would roughly double the value of $\mathcal{A}_{\text{gas}}$ compared to $E/P=100$~V/cm/bar, extrapolating from \cite{Huk}. Based on this, we set an O$_2$ concentration limit of 0.1~ppm in this work to ensure charge losses remain well below 10\%.\footnote{{Technically, the drift velocity in eq. \ref{att_Huk1} depends on impurities, too, as discussed in next sub-section. Fig. \ref{Vdrift}-right illustrates the impact of N$_2$ (that is similar to that of O$_2$ and much milder than H$_2$O). At N$_2$ (O$_2$) concentrations up to 200~ppm, as the ones employed in some of the measurements presented in \cite{Huk} and well above those discussed in this work, the dependence of $v_d$ on impurities is \% level. This extrapolates to \% level variations in the attachment coefficient. While these variations are negligible for attachment they can instead limit tracking severely, as discussed in text.}}

\subsubsection*{Modifications of the drift velocity}
\begin{figure}

    \begin{minipage}{1\textwidth}
       \centering
        \includegraphics[width=1\linewidth]{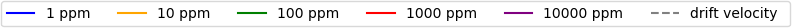}
    \end{minipage}\hfill
    \begin{minipage}{0.5\textwidth}
       \centering
        \includegraphics[width=1\linewidth]{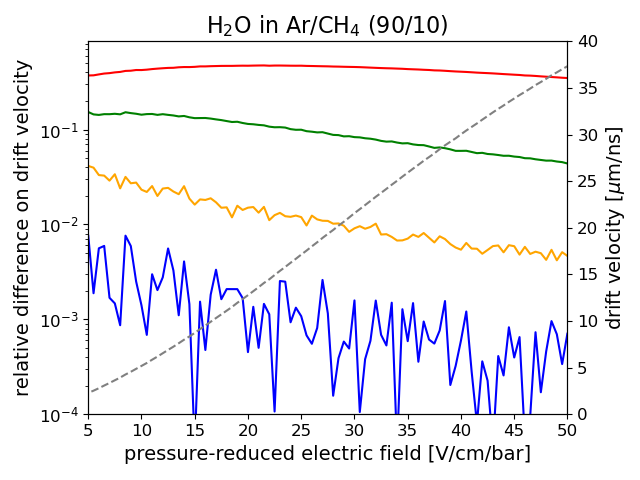}
    \end{minipage}\hfill
    \begin{minipage}{0.5\textwidth}
        \centering
        \includegraphics[width=1\linewidth]{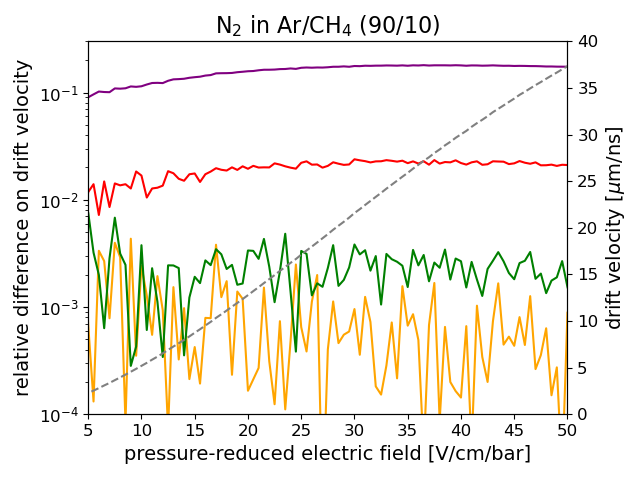}
    \end{minipage}
    \begin{minipage}{0.5\textwidth}
       \centering
        \includegraphics[width=1\linewidth]{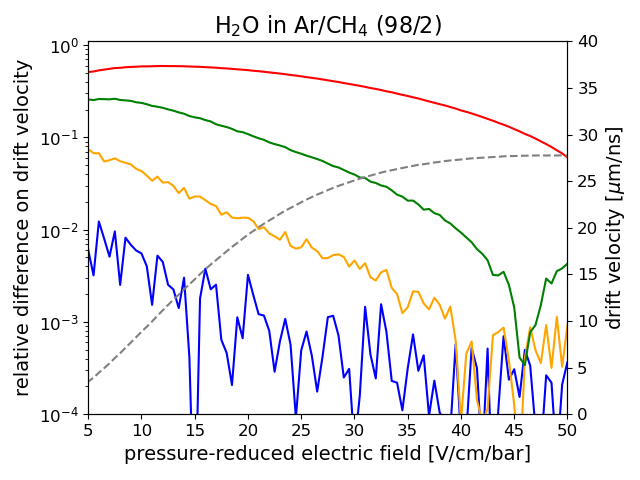}
    \end{minipage}\hfill
    \begin{minipage}{0.5\textwidth}
        \centering
        \includegraphics[width=1\linewidth]{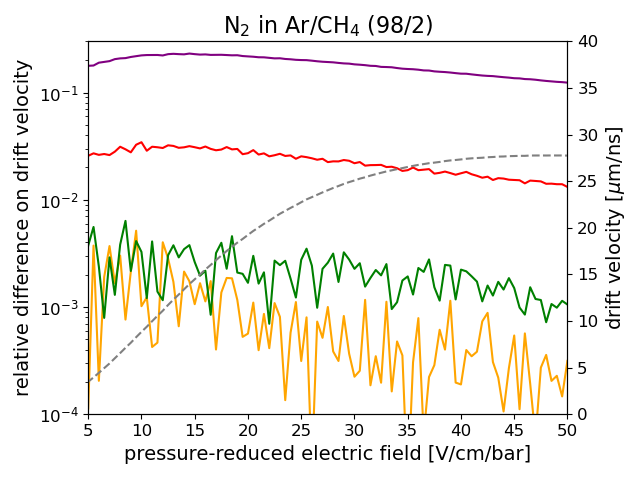}
    \end{minipage}
    \caption{Relative variation of the drift velocity ($v_d$) with respect to purified conditions, for different impurity concentrations (absolute value). Simulations have been performed with Pyboltz, focusing on Ar/CH$_4$ admixtures at $P=10~$bar and $T=25~^\circ$C. The top two figures correspond to Ar/CH$_4$ admixed at 90/10 (P10) and the bottom two to Ar/CH$_4$ at 98/2. Left: H$_2$O contamination. Right: N$_2$ contamination. Drift velocity is also shown (dashed line) with axis on the right.}
    \label{Vdrift}
\end{figure}

Impurities, even in trace-amount, can substantially modify the drift velocity to the extent of limiting particle tracking, too. Fig. \ref{Vdrift} provides the sensitivity of Ar/CH$_4$ (90/10) to H$_2$O (left) and N$_2$ (right), as a function of the pressure-reduced electric field in the range 5-50 V/cm/bar. It represents the magnitude $\left|\frac{\Delta{v_d}}{v_d}\right| = \left|\frac{v_d(f=0) ~ - ~ v_d(f)}{v_d(f=0)}\right|$, obtained with Pyboltz at $P=10~$bar and $T=25~^\circ$C.  The figure shows how, as expected, the impact of H$_2$O is stronger than that of N$_2$ by at least one order of magnitude.\footnote{ Note that achieving the sensitivity reported in the figure required using a computing cluster over few days. In case of water, this allows seeing an interesting behaviour for the 98/2 admixture, for which the $v_d$ plateau is within the range of fields explored. In this case, the variability increases below the plateau and, within it, it shows a compensating effect at specific fields (i) at 45 V/cm/bar for 100 ppm; ii) in the case of 1000 ppm, seemingly outside the region studied; iii) in the case of 10 ppm, statistical fluctuations limit the estimate, but a clear decrease is seen in the region 40-45 V/cm/bar). It is no surprise that the region of the plateau shows less sensitivity to impurities, as the impact of the cooling-effect of the H$_2$O molecule in the energy distribution of the electrons mimics effectively a change in the field, that would leave $v_d$ unchanged. However, even in the plateau, this equivalence between cooling and field only appears to apply in certain field regions. This non-trivial effect, that could find practical application if experimentally confirmed, seems to arise from the fact that the $v_d$ plateau occurs when the energy distribution of the electrons is within the steep-varying Ramsauer minimum of the elastic cross-section of the main gas. This makes the product of both (energy distribution and elastic cross-section), that is the key magnitude that enters in the evaluation of $v_d$, very sensitive to small modifications of the shape of the energy distribution.}

\begin{figure}
    \centering
    \begin{minipage}{0.5\textwidth}
       \centering
        \includegraphics[width=1\linewidth]{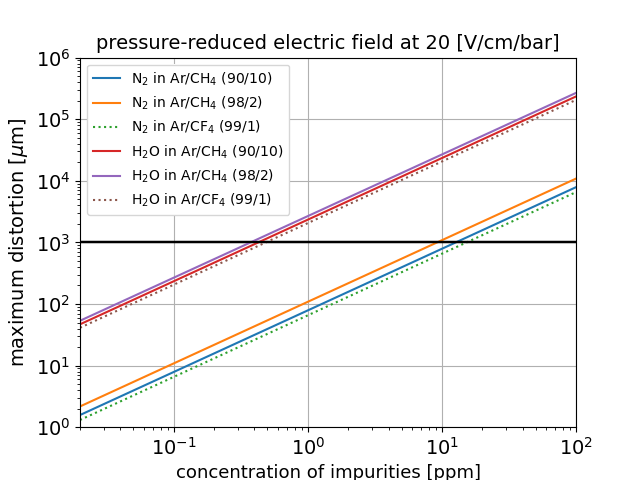}
    \end{minipage}\hfill
    \begin{minipage}{0.5\textwidth}
        \centering
        \includegraphics[width=1\linewidth]{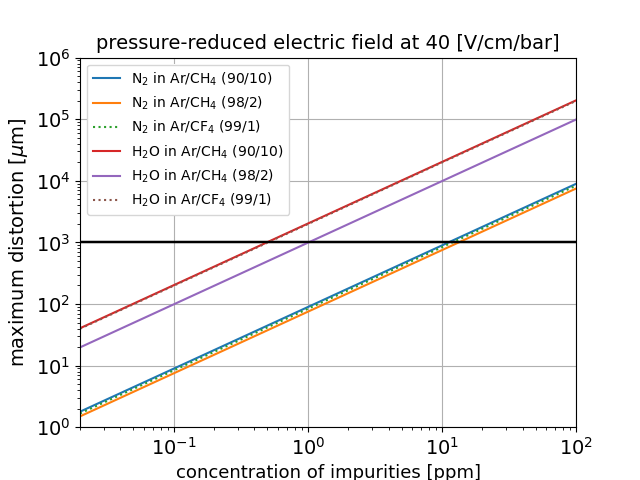}
    \end{minipage}
    \caption{Maximum track distortion obtained in simulation as a function of the concentration of contaminants in the gas medium, at 10~bar and a temperature of $25~^\circ$C. The pressure-reduced electric field is 20~V/cm/bar (left) and 40~V/cm/bar (right). Three reference gas mixtures are considered (Ar/CH$_4$ at 90/10 and 98/2, Ar/CF$_4$ at 99/1), under H$_2$O and N$_2$ contamination (the case of O$_2$ is not shown, for clarity, but it is similar to N$_2$).}
    \label{sigma_z}
\end{figure}

Deviations from the nominal (impurity-free) value of the drift velocity will cause distortions on the reconstructed tracks. We assume here a worse-case scenario where the standard calibration systems (e.g.:  laser \cite{laser}, $^{83m}$Kr source \cite{Kr}, online drift-monitor \cite{driftPhil}, cosmic muons...) either do not exist or are not able to perform the required corrections fully. In such a case, a track close to the cathode would yield a misreconstructed axial position by as much as:
\begin{equation}
\Delta{z} = L \left| \frac{\Delta v_d}{v_d} \right| \simeq \frac{L}{v_d(f=0)} \cdot {\left|\frac{dv_d}{df}\right|} \cdot f \label{vd},
\end{equation}
with $L=5$~m being the drift distance. The linear approximation on the r.h.s. of eq. \ref{vd} allows extrapolation to low concentrations ($f$), for which the numerical precision of simulations is lost. Fig. \ref{sigma_z} presents the maximal track distortion as a function of the concentration of impurities (from eq. \ref{vd}) for two typical electric fields (200~V/cm, 400~V/cm) and three gas mixtures (Ar/CH$_4$ at 90/10 and 98/2, and Ar/CF$_4$ at 99/1). For reference, the 1~mm space-resolution landmark 
is indicated with a horizontal line. As seen, depending on the system requirements, the tolerable impurity levels range from 10~ppm of N$_2$ (1~mm  distortion) down to values as low as 0.05~ppm of H$_2$O (0.1~mm distortion). The case of O$_2$ is not shown since the contamination levels that are acceptable for tracking are well above those for H$_2$O and, in most practical conditions, its presence will impact the electron lifetime much before it starts to be a concern for tracking. Besides the aforementioned limits for O$_2$, the tolerable concentrations of  N$_2$ and H$_2$O that will be assumed throughout this work will be 10~ppm and 1~ppm, respectively (limited by distortions in tracking at mm-level). They are compiled in table \ref{TabLimOut}. 

\subsubsection*{Light quenching and absorption}
Light quenching (preceding emission) and absorption (following emission) depend strongly on the gas mixture. For reference, pure Ar scintillation peaks in the vacuum ultraviolet (VUV) at around 128~nm \cite{Roberto}, and photoabsorption coefficients (at around room temperature) can reach in this region values as large as $\Pi = 100$ cm$^{-1}$ bar$^{-1}$ for common impurities \cite{ATLAS}. With light attenuation following the absorption law $\sim \exp{(-L\cdot{P}\cdot{f}\cdot\Pi)}$, a 10\% light loss on L=5~m of argon at P=10~bar would roughly arise from just 0.2~ppm of any common contaminant. In contrast, the slow decay times of the triplet component of Ar$^*_2$ ($\tau_3 = 3.2~\mu{s}$) make argon scintillation very vulnerable to deactivation transfer reactions (quenching) with impurities, too. The `triplet-dominance model' describes well the situation in conditions of interest to gaseous detectors \cite{Taka, DGD_impurities}, leading to an expression for the light loss:
\begin{equation}
   P_Q = 1 - P_{\text{scin}} \simeq  1 - \frac{1}{k \cdot P \cdot f \cdot \tau_3}. \label{quench}
\end{equation}
Here $P_{\text{scin}}$ is the scintillation probability, $k$ the quenching rate in units of s$^{-1}$ bar$^{-1}$ (room temperature assumed), and $f$ is the fraction of contaminant. Measurements of the scintillation quenching in \cite{Taka} point to 0.2~ppm as the concentration at which CO$_2$ would cause 10\%-level light losses, and about 20~ppm for N$_2$ (at 10~bar pressure).

\begin{table}[]
  \centering    
\begin{tabular}{ |c||c|c|  }
\hline
Species & Limit value & Limiting factor \\
\hline
\hline
O$_2$ & $\sim$ 0.1 ppm & Attachment \\  \hline
H$_2$O & $\sim$ 1 ppm & Drift velocity \\  \hline
N$_2$  & $\sim$ 10 ppm & Drift velocity \\  \hline
\end{tabular}
\caption{Limiting values for each contaminant.}
\label{TabLimOut}
\end{table}

The experimental values obtained above, for the maximum concentrations of impurities that are compatible with argon scintillation, are at the same level (or even below) than those obtained from the analysis of the charge properties in regard to attachment and drift velocity. 
The situation for Ar/CF$_4$ (99/1) is expected to be much more favourable, though. First, about 90\% of the usable scintillation lies in the near-UV and visible range (200-800~nm) in that case \cite{Pansky, Amedo_spe}. Considering the absorption coefficient for H$_2$O at the low-end of such wavelength range, a value of $\Pi = 10^{-2}$ cm$^{-1}$ bar$^{-1}$ is expected \cite{ATLAS}. As a result, 10\%-level light losses would require of H$_2$O concentrations as high as 2000~ppm (being water the most opaque contaminant, a priori).
Regarding light quenching, it has been measured in \cite{Margato} for pure CF$_4$ at 1~bar to be 50\% under a N$_2$ contamination of 4\%. Assuming a quenching relation of the type \ref{quench}, 4000~ppm of N$_2$ would be needed to cause a 10\% light loss. The factor 200 above the values obtained for pure argon matches well the expectation in eq. \ref{quench} resulting from the much shorter lifetime of CF$_4$ states, in the range 10-20~ns \cite{CF4Yields}. Following the analogy with the argon case further, a plausible estimate of the maximum amount of (more reactive) molecules such as CO$_2$ and H$_2$O would be 40~ppm, greatly exceeding the upper bounds obtained from charged transport. 


\subsection{Outgassing model} \label{outgas}

Outgassing is dominated by the technical plastics employed in TPC construction, that is orders of magnitude above that of metals, e.g., from the TPC vessel or field cage. We consider Teflon/PTFE (commonly used as a light reflector and as a structural material for field cages \cite{XENONtef,NEXTtef}), PMMA (a sturdy material with good optical properties) and PEEK (a low-outgassing structural material). Plastics follow, in general, an outgassing law depending on time ($t$) as $Q(t) \sim t^{-a}$ with $a\sim 1/2$, corresponding to a diffusion-driven bulk desorption law. This is in principle applicable as long as i) the desorbed gas remains far from the saturation concentration of the exterior (about $f=3000$~ppm in case of H$_2$O at T=25$~^\circ$C, P=10~bar) and ii) the material is thick enough to keep furnishing the contaminant at any given time. As time passes, however, the desorbed gas reaches out from deeper regions until the material boundaries are approached, moment at which outgassing drops exponentially. The process leads to the following piece-wise outgassing law \cite{DiffLaw}:
\begin{align}
    Q(t) = \begin{cases} 
    k \sqrt{\frac{\pi \tau}{t}} \quad ~~~~\text{for } t \leq 0.5 \tau, \\
    4\, k \, e^{-t/\tau } \quad \text{for } t > 0.5 \tau. \\
    \end{cases}
    \label{outgasLaw}
\end{align}
{
Here, $k$ indicates the strength of the outgassing rate, and $\tau$ is the time constant that characterizes the moment at which the depth-of-origin of the desorbed species approaches the material thickness $t_m$. 
They have a dependence with temperature since both depend on the coefficient of diffusion $D_{bulk}$, which is given by
\begin{equation*}
    D_\text{bulk} = D_\text{bulk,0} \cdot \exp{\left(-\frac{E^*}{k\,T}\right)},
\end{equation*}
where $E^*$ represents the activation energy, $k$ is the Boltzmann constant, and $T$ is the temperature.}
The time constant $\tau$ is computed as $\tau= \frac{t_m^2}{\pi^2 D_{\text{bulk}}}$. A typical value for technical plastics, obtained in an outgassing setup, is $D_{\text{bulk}} = (3.0 \pm 1.2) \times 10^{-9}$ cm$^2$ s$^{-1}$ \cite{CERN_master}.\footnote{{ Diffusion coefficients, $D_{\text{bulk}}$, larger by up to $1000$ times and $10$ times can be found through direct measurements (e.g., \cite{Diffusion}) for Teflon and PMMA, respectively. Similarly, the low-concentration diffusion coefficient of N$_2$ in PEEK \cite{Diffusion2} is about $10$ times larger than the one used in text. Such high diffusion coefficients, if applicable to outgassing, would result in Teflon being depleted within $\tau \sim \text{minutes}$ (eq. \ref{outgasLaw}-bottom), which contradicts common experience.  The diffusion coefficient assumed in this work, however, does not cause depletion of the outgassing species over the time scales discussed (several days), when considering practical thicknesses above 2~mm. The potential presence of depletion effects at time scales shorter than the ones considered here represents an optimistic scenario, that needs to be addressed experimentally, and whose discussion we opted for leaving out of this work.}} 

Outgassing rates are typically measured experimentally in vacuum and can be parameterized through the value obtained 10 hours after pumping ($Q_{\text{10h}}$) and the exponent $a$ \cite{Poole} (the latter assumed to be 0.5 hereafter). For instance, in the case of Teflon at room temperature, $Q_{\text{10h}}$ typically ranges from $3.5\times 10^{-8}$ to $1\times10^{-7}$ $\text{mbar}\cdot\text{l}\cdot\text{cm}^{-2}\cdot\text{s}^{-1}$ (for a recent review, see \cite{OG2}). Mass spectroscopy indicates that the O$_2$ content in outgassing is generally less than 10\% of the total, while N$_2$ and H$_2$O concentrations vary considerably, ranging from nearly 70\% H$_2$O and 30\% N$_2$/CO to almost 100\% N$_2$ and 0\% H$_2$O \cite{Thieme, Poole, Joshi}. For both H$_2$O and N$_2$, we consider the most unfavourable case (i.e., that they constitute 100\% of the outgassing), together with the middle (`typical') and lowest (`optimistic') $Q_{\text{10h}}$ values reported for total outgassing rates in \cite{OG2}. For O$_2$, we consider a range between the value directly measured in \cite{Thieme} (`optimistic') and 10\% of the middle value of $Q_{\text{10h}}$ reported in \cite{OG2} (`typical'). Since outgassing rates depend on material quality and surface finish, our approach has been to select a range that includes both typical outgassing values and the lowest ones found in the literature. Measurements in \cite{Thieme} are particularly relevant as they do not involve bake-out, thereby matching the expected experimental conditions under which large-volume TPCs are commissioned.

Outgassing rates for PMMA and PEEK are found in the range $Q_{\text{10h}}=10^{-7}$--$10^{-6}$ $\frac{\textnormal{mbar}\cdot{\textnormal{l}}}{\textnormal{cm}^2\cdot{\textnormal{s}}}$. For PMMA, these rates tend to be dominated by H$_2$O, with no trace of O$_2$ or N$_2$ being detected (e.g., \cite{Thieme}). For PEEK, the situation is similar; however, the amount of O$_2$ reported in the literature is at the \%-level (e.g., \cite{CERN2nd}). As PMMA is typically used as a window for light detection \cite{DarkSide} and PEEK is sometimes present in readout frames \cite{ARIADNE}, one can anticipate that the use of these materials would be restricted to the end-caps, constituting at most 20\% of the total outgassing surface in our default geometry. Thus, the contribution of PMMA and PEEK to outgassing can generally be made subdominant in the presence of a Teflon reflector, if these materials are selected from the lower end of the range of measured values. 
{ A compilation of the outgassing rates considered in this work can be found in table \ref{OutTab} (column 2).}

In the following, we will adopt the outgassing law in eq.~\ref{outgasLaw} for studying the impact of H$_2$O and N$_2$ on TPC performance. For O$_2$ (a priori the most critical case), we will perform the calculations assuming a more extreme scenario in which the outgassing remains constant at $Q=Q_{\text{10h}}$. In all cases, Teflon will be assumed to dominate the system outgassing.

\begin{table}[]
  \centering    
\begin{tabular}{ |c||c|c|c|c|c|  }
\hline
species & Q$_{10h}$ [$\frac{\textnormal{mbar}\cdot{\textnormal{l}}}{\textnormal{cm}^2\cdot{\textnormal{s}}}$] & model & ppm & D$_\text{x,bulk}$ [cm$^2$/s] & D$_\text{x,He/Ne/Ar/Xe}$ [cm$^2$/s]\\
 \hline \hline
 O$_2$  & $(0.05-1)\times10^{-8}$   &  const.  & $\sim 0.1$ &  $(3 \pm 1.2) \times 10^{-9}$    & 0.0725~/~0.0316~/~0.0195~/~0.0126\\
 \hline
 H$_2$O & $ (0.35-1) \times 10^{-7} $   &  $t^{-1/2}$  & $\sim 1$   &   $(3 \pm 1.2) \times 10^{-9}$   & 0.084~/~0.0397~/~0.0252~/~0.0173\\
 \hline
 N$_2$  & $ (0.35-1) \times 10^{-7} $   &  $t^{-1/2}$  & $\sim 10$  &   $(3 \pm 1.2) \times 10^{-9}$   & 0.0692~/~0.0310~/~0.0194~/~0.0128\\
 \hline
\end{tabular}
   \caption{Lower (`optimistic') and upper (`typical') limits to the outgassing rates considered in this work, for different gas species. A $Q \sim Q_{10\text{h}} \cdot \sqrt{t/10h}$ law has been assumed for H$_2$O and N$_2$, and a conservative $Q \sim Q_{10\text{h}}$ law for O$_2$ (third column). The fourth column shows the typical limiting concentrations for TPC operation under argon gas at 10~bar, as obtained in text. The fifth column provides an estimate of the effective diffusion constant of the outgassed species in typical technical plastics. Finally, the last column compiles the diffusion constants of common outgassed species (x $=$ O$_2$, H$_2$O, N$_2$) in helium, neon, argon and xenon gas at $P=10$~bar and T$=20~^\circ$C.}
    \label{OutTab}
\end{table}

Once the outgassed species reach the material surface, further diffusion into the exterior volume will be suppressed depending on the gas pressure. The diffusion coefficient of species `x' (x$\equiv$ O$_2$, H$_2$O, N$_2$) in a given gas (helium, neon, argon and xenon will be considered here) can be evaluated following a formula based on Fuller's work \cite{Fuller, FullerBook} as:
\begin{equation}
D_\text{x, He/Ne/Ar/Xe}=\frac{0.00143\cdot T^{1.75}}{P\sqrt{\frac{2}{\frac{1}{M_\text{x}}+\frac{1}{M_\text{He/Ne/Ar/Xe}}}}\cdot[\sqrt[3]{V_\text{x}}+\sqrt[3]{V_\text{He/Ne/Ar/Xe}}]^2},
\end{equation}
in units of cm$^2$/s. Here $M$ is the molecular mass of each gas, $P$ is the pressure in bar, $T$ is the temperature in K, and $V_i$ are non-dimensional `experimental diffusion volumes'. 
The expression above can be evaluated for admixtures ($B_1$ + $B_2$), as long as contaminants remain at concentrations much below those of the gas constituents:
\begin{equation}
\frac{P}{D_{\text{x}, B_1+B_2}}=\frac{P_{B_1}}{D_{\text{x}, B_1}} + \frac{P_{B_2}}{D_{\text{x}, B_2}}.
\end{equation}
In the above equations, $P_i$ represent the partial pressures of each gas in the mixture. Diffusion volumes can be extracted from \cite{FullerBook}: $V_\text{Ar}= 16.2$, $V_{\text{H}_2\text{O}}= 13.1$, $V_{\text{N}_2}= 18.5$, $V_{\text{O}_2}= 16.3$, $V_{\text{CH}_4} \simeq V_C + 4V_H= 25.1$, $V_\text{Xe}= 32.7$, $V_\text{He}= 2.67$, $V_\text{Ne}= 5.98$ and $V_{\text{CF}_4} \simeq V_C + 4V_F= 74.7$. For the admixtures discussed in this work, the diffusion coefficients are sufficiently close to those in the main gas so that we may neglect the contribution from the molecular additive (a compilation can be found in table \ref{OutTab}, last column).

It should be noted that additional mechanisms at the material-gas interface could, in principle, introduce a pressure dependence and an outgassing law different from the one assumed in vacuum. This would need to be elucidated through direct measurements. As with other assumptions employed earlier in the text, the aim of our study is to provide upper limits to the impact of impurities in an actual system and verify that they can be kept below potentially harmful levels through reasonable design choices. In this regard, there is no obvious mechanism by which outgassing could be enhanced as pressure increases, compared to the model proposed here.

\subsection{Gas distribution scheme}\label{GasScheme}

\begin{figure}[h!!!]
    \centering
    \includegraphics[width=0.9\textwidth]{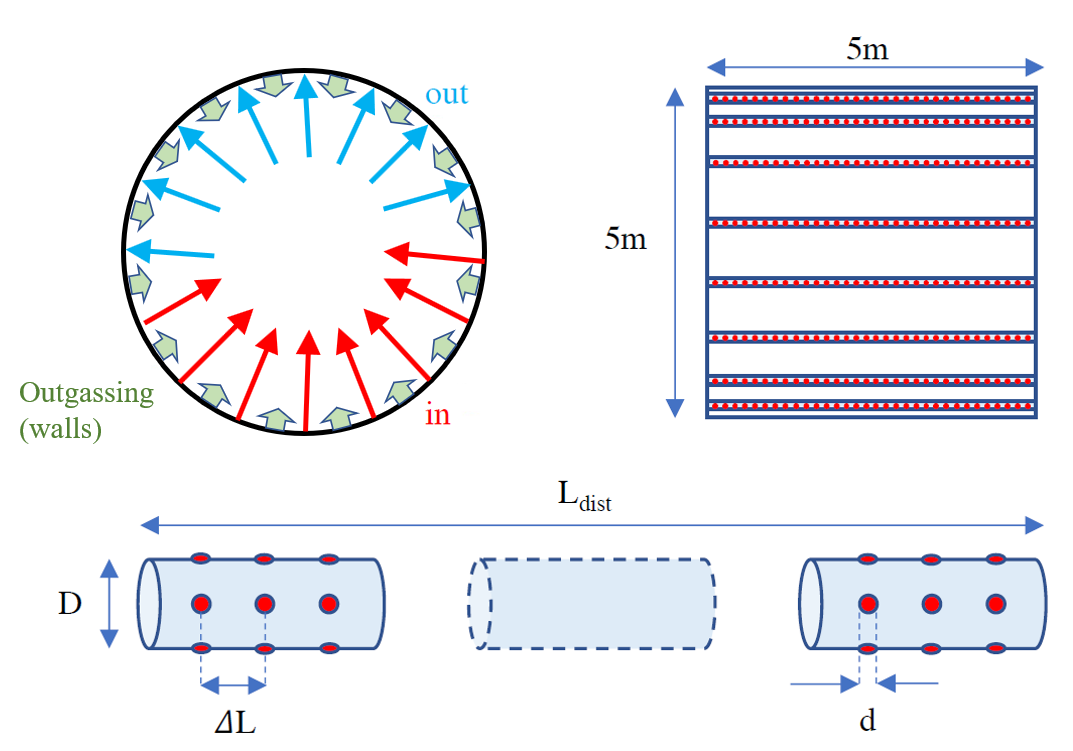}
    \caption{Top: cross-section of the proposed gas distribution inside the field cage (left), with perforated distribution rods extending axially along it (right). Eight injection rods are placed at the bottom and eight ejection rods at the top. The field cage is assumed to be gas-tight, extending along 5~m in length and 5~m in diameter. Holes are indicated by circles (not to scale). Bottom: close-up of a gas distributor rod.} \label{Ref_Fig}
\end{figure}

Our baseline proposal for the gas distribution is based on axial rods (perforated plastic tubes) attached to the field-cage, with inlets positioned at the bottom-half and outlets at the top-half, as shown in Fig.~\ref{Ref_Fig} (top-left). The rationale for this design is to enable a straightforward implementation, as the field-cage is generally the least instrumented part of a TPC. This concept does not interfere with the end-cap design and, with appropriate choices, can ensure uniform gas distribution along the drift/axial direction. Sixteen equispaced axial rods run along the rectangular section of the TPC (eight for injection and eight for extraction) as shown in Fig.~\ref{Ref_Fig} (top-right), providing an inter-rod distance of nearly 1~m between centres. As demonstrated later, this arrangement is sufficient to efficiently sweep outgassing from the field-cage walls. The distributor rods are periodically perforated with a pattern that is discussed in the next section.

{Assuming that gas is continuously recirculated and that clean gas is injected (as discussed previously)}, and in the absence of gas leaks, the ultimate purity levels that can be achieved are limited by the capacity of the recirculation compressors. For reference, the ALICE TPC circulates gas at 20~m$^3$/h (1~bar) while the NEXT-100 experiment is designed to circulate at approximately 5~m$^3$/h (10~bar), \cite{MonraPrivate}. Based on this, we will consider two different scenarios for a 10~bar system, referring to them as `typical' (30\% above NEXT flow: 6.5~m$^3$/h) and `optimistic' (20~m$^3$/h), as summarised later in table \ref{DistTab}.\footnote{We will not refer to normal litres in text. All units of volume must be understood at the given pressure.}

\section{Fluid dynamics model} \label{CFD}

We discuss, separately: i) the optimization of the geometry of the distributor rods, performed through a parametric model based on generalized continuity equations and ii) the CFD simulations of the proposed gas distribution scheme, focusing on the concentrations of contaminants in space and time, performed with COMSOL Multiphysics v5.5 \cite{COMSOL} (`Fluid Flow' Module). 

\subsection{Optimization of the distributor rods} \label{sectionRods}

A qualitative discussion of the main features of gas distributors has been left to appendix (section \ref{iterative_script}). Two reference cases are considered: a capped distributor and an open one. In general grounds, an open distributor is less efficient, given that only a fraction of the gas exits through the holes and into the chamber --as little as a few \% for high gas flow. 
It also leads to a higher pressure drop along the distributor due to friction, significantly greater directionality of the outgoing gas and a non-uniform discharge pattern. Low efficiency in combination with a directional and non-uniform discharge pattern will adversely affect the distribution of impurities throughout the chamber and the time required to achieve uniform mixing with the main gas. Therefore, we opted for a capped distributor, ensuring uniform and normal injection of all the circulated gas.

A large number of holes per gas distributor, on the order of several hundreds, has been proposed in previous large-volume TPCs (e.g., \cite{ALICE_TPC}).
Optimising such a geometry, with each hole potentially down to $\mathcal{O}($mm) in size and each distributor rod extending over a length of $5~$m, is impractical when using numerical simulations. As an alternative, we opted for a parametric approach based on the balance equations for each hole section, which allows the solution to be obtained iteratively. We followed the classical formulation of \cite{Bailey}, whereby: i) a discharge coefficient $C_d$ (between 0 and 1) regulates the loss of flow into the distributor holes, \thinspace--- mass balance; ii) a regain coefficient $C_r$ (between 0 and 2) regulates the pressure increase along the distributor after the ensuing velocity drop, \thinspace--- momentum balance (Fig.~\ref{COMSOL_1D}-right, section \ref{iterative_script} of appendix); iii) an angular coefficient $\gamma$ regulates the angle of discharge relative to the normal:
\begin{eqnarray}
    D^2(v_\text{u} - v_\text{d}) &=& d^2C_d\sqrt{\frac{\Delta{P_\text{hole}}}{\rho/2}}, \label{Rec1}\\
    P_\text{d} - P_\text{u} &=& C_r\frac{1}{2}\rho(v_\text{u}^2 - v_\text{d}^2), \label{Rec2}\\
    \theta &=& \gamma \cdot \arctan\left(\sqrt{\frac{\frac{1}{2}\rho v_\text{u}^2}{\Delta{P}_\text{hole}}}\right) .\label{Rec3}
\end{eqnarray}
In the above expressions $D$ refers to the diameter of the distributor, $d$ is the hole diameter and $\rho$ the gas density, whereas sub-indexes `up' and `down' refer to upstream and downstream of a given hole. Velocities ($v$) and pressures ($P$) must be understood as averages over the distributor cross-section. $\Delta{P_\text{hole}}$ is the pressure difference between the distributor and the TPC at the hole position, that drives the hole discharge:
\begin{equation}
\Delta{P_\text{hole}} = \frac{P_\text{u}+P_\text{d}}{2}-P_{_\text{TPC}}.
\end{equation}
Losses due to friction will occur when the gas transits between adjacent holes; however, in the conditions discussed here, the effect is negligible. For each distributor topology, three main quantities have been evaluated through the iterative solution of the balance equations: the pressure difference between the distributor and the TPC at the position of each hole ($\Delta{P_{\text{hole}}}$), the velocity of the gas as it flows through it ($v_{\text{hole}}$), and its maximum discharge angle relative to the normal ($\theta_{\text{max}}$).

In the conditions studied here, as $d\ll{D}$ and the discharge velocity is both small and very similar for all holes, it is possible to approximate the above magnitudes by (see appendix, {section~\ref{analytic})}):
\begin{eqnarray}
    \Delta{P}_\text{hole} & \simeq & \frac{1}{2} \frac{1}{C_d^2} \rho v_\text{hole}^2, \label{ana_1}\\    
    v_\text{hole} & \simeq & \frac{\Phi}{\textnormal{t.h.a.}}, \label{ana_2}\\
   \theta_\text{max} & \simeq & \gamma C_d \frac{d^2}{D^2} \frac{L_\text{dist}}{\Delta{L}}, \label{ana_last}
\end{eqnarray}
with $\Phi$ being the total gas flow injected into the TPC, t.h.a. being the total hole area, $L_\text{dist}$ the length of the distributor, $\Delta{L}$ the distance between holes, and the total hole area relates to the number of distributors and holes per distributor as:
\begin{equation}
   \text{t.h.a.} = N_\text{dist} \cdot N_\text{holes/dist}\cdot \frac{\pi}{4} d^2.
\end{equation}

In order to come to some plausible distributor geometries, the following design constraints were introduced: i) a diameter hole not much smaller than 1~mm, ii) a diameter of the distributor not much larger than 5~cm, iii) an inter-hole distance not larger than 5~cm, iv) a pressure drop below 10~mbar, v) a discharge velocity well below the speed of sound, and vi) a maximum discharge angle below or around $5^\circ$. 
Based on these requirements, the main challenge is to obtain a small discharge angle without neither letting the inter-hole distance and tube diameter grow too big, nor the hole diameter become too small. Four exemplary geometries satisfying the above conditions ($A$-$D$) are compiled in table \ref{DistTab}, 
{
although many more combinations are obviously possible.}

{
It might be argued that small distributor diameters like those given in $C$ and $D$ should be preferred, so as not to reduce the fiducial volume due to sheer penetration in the active volume or charging-up.
However, for hole velocities above 1-2~m/s (the velocity present in geometries $C$ and $D$) we observed a tendency of the flow to form large-scale vortices.
Those vortices substantially delayed the time for impurities to achieve uniform mixing. Aiming at a comparative study, we opted for a design that guaranteed a range of velocities in the default distributor geometry outside the region where the effect appears, for the gas flows studied.
Furthermore, even for geometries $A$ and $B$, small distributor diameters can also be used through minor adjustments of the geometry or increasing the thickness of the distributor walls (studies that fall outside the scope of this work)\footnote{For instance, a reduction from a diameter of 7~cm to 3.5~cm in geometry $B$ can be achieved by increasing the number of injection rods to 32 and increasing the inter-hole distance to 60~mm, yet keeping the rest of the fluid properties as in table \ref{DistTab}.}.
Therefore, we have focused in the remainder of the text on $A$ (single hole) and $B$ (triple hole), taking the latter as the default distributor geometry. We wish to recall the general wisdom that field distortions in a field cage decrease quickly when the distance to the feature (distributor, in this case) extends by $\times 2$ its size, and uniform charging-up (a plausible assumption under uniform irradiation
or under the use of electrostatic-dissipation transparent coatings \cite{Kentaro}) will theoretically reduce the extent of field distortions to zero. As a consequence, it is expected that, even in a worst-case scenario, a generous fiducialisation of up to 10-20~cm (radially inwards) will suppress most (if not all) of the effect stemming from charging-up of the Teflon and distributors, in geometries $B$ and $D$. We plan to discuss this as part of a separate paper, but details are too context-dependent and out of scope.}

\begin{figure} [h!!]
    \centering
    \begin{minipage}{1\textwidth}
       \centering
        \includegraphics[width=1\linewidth]{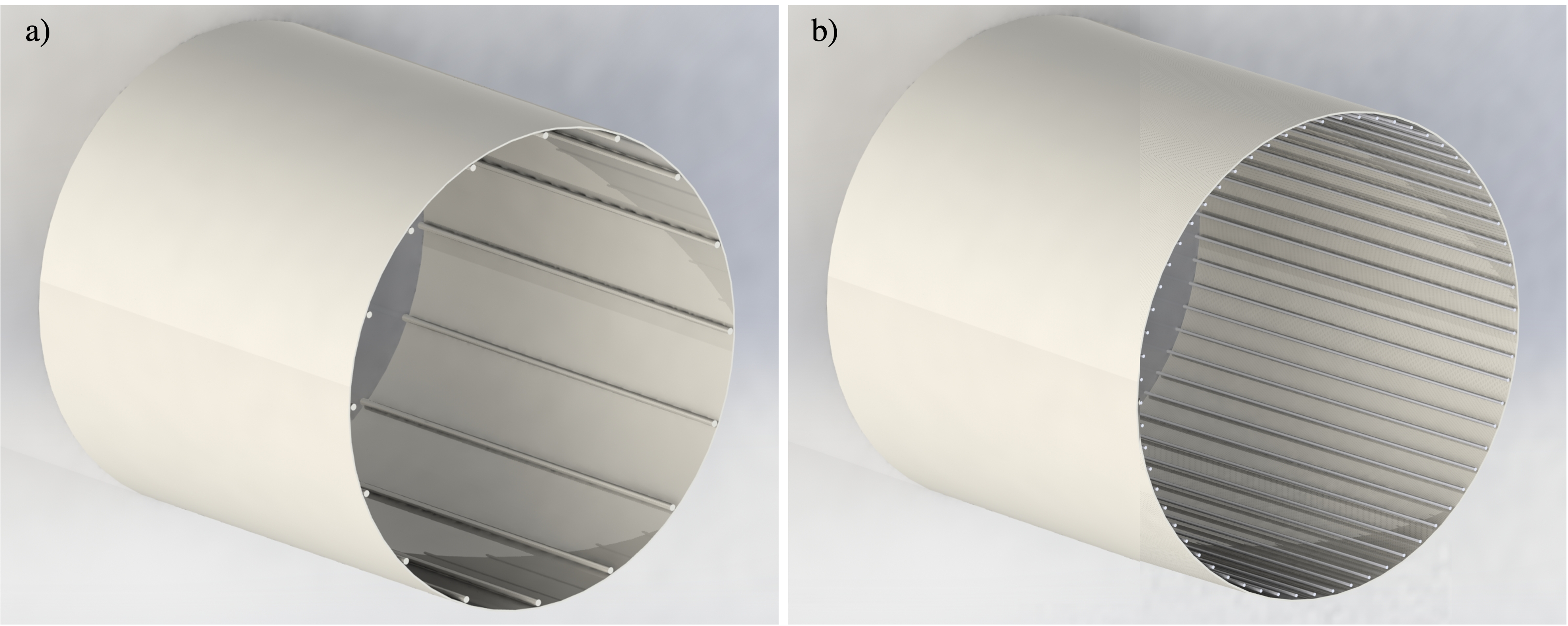}
    \end{minipage}\hfill
    \caption{Left (a): gas distribution corresponding to geometry $B$ in table \ref{DistTab}, the default geometry studied in text. Right (b): modification of geometry $B$, employing $\times 4$ tubes with $\times 2$ smaller diameter, and holes separated $\times 4$, thus matching the same initial conditions of the fluid at the inlets. \label{Tef_figs}}
\end{figure}

A CAD image corresponding to geometry $B$ is shown in Fig. \ref{Tef_figs}-a, whereas Fig. \ref{Tef_figs}-b provides an equivalent geometry in terms of fluid parameters at the inlets, however with smaller-diameter distributors.

\begin{table}[h!!!]
  \centering    
\begin{tabular}{ |c||c|c|c|c|c|c|c|c|  }
\hline
flow [m$^3$/h] & name & \# holes & $\Delta{L}$ [mm] & D[mm] & d[mm] & $\overline{{\Delta}P_h}$[mbar] & $\overline{v_h}$[m/s] & $\theta_\text{max}$ [deg] \\
 \hline \hline
6.5 & A & 1 & 15 & 45 & 1 & 0.15 & 0.865 & 4.2\\
 \hline
20  & A & 1 & 15 & 45 & 1 & 1.46 & 2.66 & 4.2\\
 \hline
6.5 & B & 3 & 15 &  70  & 1  & 0.017   & 0.29 & 5.1\\
 \hline
20  & B & 3 & 15 &  70  & 1  & 0.16   & 0.88 & 5.1\\
 \hline
6.5 & C & 1 & 15 &  30  & 0.75  & 0.49   & 1.54 & 5.3\\
 \hline
20  & C & 1 & 15 &  30  & 0.75  & 5.78   & 5.29 & 5.3\\
 \hline 
6.5 & D & 3 & 50 &  30  & 0.75  & 0.61   & 1.72 & 4.7\\
 \hline   
20  & D & 3 & 50 &  30  & 0.75  & 4.63   & 4.73 & 4.7\\
 \hline    
\end{tabular}
   \caption{Characteristics of the gas-distributor geometries discussed in this work. The first column indicates the gas flow (assumed to be Ar/CF$_4$ (99/1) at 10~bar and $25~^\circ$C). Columns 4-6 indicate the geometrical parameters of the distributor. Columns 7-9 indicate the average value of the pressure drop relative to the TPC, average discharge velocity, and maximum discharge angle of emission, respectively. Eight distributor rods of 5~m in length were considered in all cases, for the injection. Geometries A (single hole) and B (triple hole) are used as default for the CFD simulation, while geometries C (single hole) and D (triple hole) aim at illustrating that it is possible to reduce the diameter of the distributor rod, keeping a low value of the discharge angle. Column 3 indicates whether the perforation pattern is made out of individual holes or a triplet (one hole placed radially and the other two azimuthally).}
    \label{DistTab}
\end{table}

\begin{figure}[h!!!]
    \centering
    \includegraphics[scale=0.55]{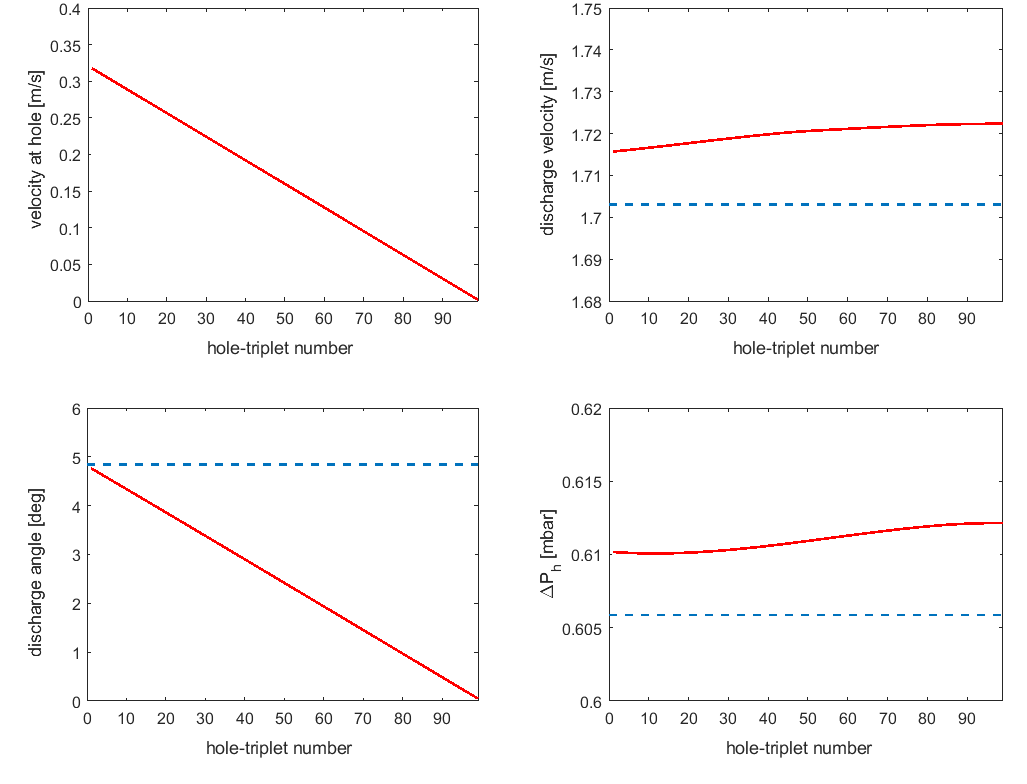}
    \caption{Main magnitudes characterizing the response of the distributor rod for each triple-hole along the distributor, shown with red lines (the first triple-hole upstream is number 1). We consider an argon gas flow of 6.5~m$^3$/h at 10~bar, 8 distributors of 3~cm in diameter and 5~m in length. The hole size is 0.75~mm and holes are separated by 5~cm (geometry $D$ in table \ref{DistTab}). Top-left: gas velocity in the distributor at the hole position. Top-right: discharge velocity. Bottom-left: angle relative to normal. Bottom-right: pressure difference relative to the TPC. (The analytical expressions given in text are shown as dashed lines).} 
    \label{1Dscript}
\end{figure}

{
An example of the results obtained is presented in Fig. \ref{1Dscript} (for details on the numerical procedure, the reader is referred to the appendix --section \ref{iterative_script}). 
The analytical estimates detailed previously are included as blue dashed lines, showing excellent agreement with the numerical solution. It should be noted that, with independence from the distributor geometry, $C_d$ approaches a constant value around $C_d = 0.61$-$0.64$ when $\Delta{P}_\text{hole} \gg \frac{1}{2}\rho v_\text{dist}^2$ \cite{Hegge}. A detailed comparison with experimental data is presented in \cite{Bailey}, providing an estimate of $C_d = 0.63$ (which is adopted hereafter). Such pressure-dominated conditions are typical of capped distributors and were verified for all our study cases. Similarly, the $\gamma$ pre-factor is largely independent of the gas flow characteristics and, according to \cite{Bailey}, weakly dependent on the geometry, and so the value is set to 
$\gamma = 0.71$. The least constrained coefficient, $C_r$, as well as the friction losses, are immaterial in the conditions discussed, as is made clear through the analytical expressions in \ref{ana_1}-\ref{ana_last}. The aforementioned values for the parameters $\gamma$ and $C_d$ are consistent with the CFD analysis of the control geometry displayed in Fig. \ref{COMSOL_1D} (appendix).
}

\subsection{CFD simulations} \label{CFD_simuls}

A numerical evaluation, based on the well-established CFD methodology, allows for a realistic approach to describe the gas and impurity dynamics by calculating the velocity field in the overall volume of the field cage, including the injection setup. It is assumed that impurities, present in trace amounts, do not modify the velocity field of the main gas. 
Translational invariance along the (axial) $z$ dimension $(\frac{\partial}{\partial z}=0)$ was assumed. This assumption relies on the (axial) orientation of the distributors and the discharge optimisation performed in the previous section, which ensures the discharge angle is nearly perpendicular to the distributor (table \ref{DistTab}). As discussed, the outgassing rate from the end-caps can be made sub-dominant relative to that from the field cage, approximately preserving the adopted symmetry. Although the region between holes cannot be well described in this approach, the distance chosen in geometries $A$-$D$ in table \ref{DistTab} is sufficiently small (15-50~mm) to anticipate a negligible (local, at most) impact.

The hole aperture of the inlets and local pressure difference (distributor-TPC) were set to provide (simultaneously) the desired average velocity per hole and flow per unit length, based on the values in table \ref{DistTab}. Achieving convergence in the simulations by setting the hole velocity instead of the pressure difference as an initial condition was not generally possible. The outlets were set to a fixed pressure of 10~bar. Gravity was included in the simulation, and in particular the hydrostatic pressure was automatically added at the inlets and outlets depending on their height.
Outgassing was set perpendicular to the walls and to the distributors, as shown in Fig. \ref{Ref_Fig}.
{
At the surface boundaries, the `no slip' condition was used \cite{NoSlip2}, as is customary for solid boundaries and Newtonian fluids in macro-scale flow and non-rarefied conditions.
The fluid properties used were those in the default COMSOL database; see \cite{Comsol_mesh} pp.~765-766 for details.}

{
Natural convection in the TPC gas has been considered in the design, although it was not explicitly included in the simulation due to its limited impact under the specific operating conditions. 
While temperature gradients can induce convective circulation, the flow regime in the system is primarily dominated by controlled gas injection and forced recirculation, which minimises the influence of natural convection on impurity redistribution. 
To assess its impact, estimates based on the Rayleigh number indicate that, given the system's dimensions and thermal conditions, convective patterns would develop on timescales significantly longer than those observed in our CFD simulations. 
Moreover, the thermal stability of the system, together with the encapsulation of heat-generating components and the low thermal conductivity of the surrounding materials, significantly reduces the magnitude of the thermal gradients responsible for convection. Therefore, the isothermal approach adopted in the model remains appropriate within the scope of the present simulation.
}

{
The mesh is composed of 701 thousand linear (Lagrange P1) elements with an average skewness of 0.9 and average growth rate of 0.87 (in both metrics 1 means a perfect element and 0 a degenerated one, see \cite{Comsol_mesh} p.~629). 
The mesh is finer closer to the holes and coarser towards the centre of the field cage. 
To ensure convergence, the solution was checked to be stable upon global up- and down-scaling of the mesh size and variations in the time step.
}

The turbulent-flow `k-$\varepsilon$' model was selected, being adequate for simulating single-phase flows at high Reynolds numbers and suitable for incompressible or weakly compressible flows, and compressible flows at low Mach numbers. The `k-$\varepsilon$' model is the most validated turbulence model and performs well in confined flows where the Reynolds shear stresses are important.
This model uses the Reynolds-averaged Navier-Stokes equations (RANS), introducing two additional magnitudes: the turbulent energy ($k$) and the rate of turbulent energy dissipation ($\varepsilon$). 
These magnitudes have their corresponding differential equations
coupling the velocity field ($\overrightarrow{v}$) with the viscous stress tensor {\textbf{K}}; 
{
the details can be found in \cite{Comsol_turbulence}.}
The dimensionless coefficients, called closure coefficients, of the model are \cite{Launder}: 
\begin{equation*}
C_{\varepsilon1}=1.44, \ \ C_{\varepsilon2}=1.92, \ \ C_{\mu}=0.09, \ \  \sigma_k=1, \ \  \sigma_\varepsilon =1.3.
\end{equation*}


\noindent  Finally, the transport of impurities (trace amount) is implemented based on Fick's law: 
\begin{equation}
    \overrightarrow{J}=-D\overrightarrow{\nabla}c,
\end{equation}
where $\overrightarrow{J}$ is the particle flow per unit area and time, and $c$ the impurity concentration in [L$^{-3}$]. This equation must be coupled to the continuity equation in presence of convection:
\begin{equation}
\frac{\partial c}{\partial t}+\overrightarrow{\nabla}\cdot\overrightarrow{J}+\overrightarrow{v}\cdot\overrightarrow{\nabla}c=0,
\end{equation}
and diffusion coefficients are taken from the last column in table \ref{OutTab}. 

{
As a summary, the following boundary conditions and models were used:
\begin{itemize}
    \item Translational invariance (2D simulation).
    \item TPC initial pressure: 10~bar ($20^{\circ}$C).
    \item Gas outlets: 10 bar (corrected by hydrostatics).
    \item Gas inlets: pressure adjusted to provide the desired gas velocity at injection, as estimated from the 1D distributor model.  The size of the hole (effectively a slit in a 2D simulation) is adjusted to provide the desired mass flow per unit length, as estimated from the 1D distributor model.
    \item No slip at the walls.
    \item Outgassing source perpendicular to the walls.
    \item $k-\varepsilon$ model for the turbulence.
    \item Fick's law for the transport of impurities.
\end{itemize}

}

\section{Results} \label{results}

\subsection{TPC filling} 

A priori, the displacement of air during the chamber filling can be performed at flow rates higher than those imposed by the recirculation compressors, if need be. In the ideal case, it might even be possible to evacuate the TPC, although for large volumes this cannot be taken for granted. Here we consider, for illustration, the process of TPC filling at a flow rate of 6.5~m$^3$/h (`typical' flow), in the default geometry ($B$). It has been assumed, for simplicity, that the chamber is filled quickly (and uniformly) up to 9~bar of Ar and 1~bar of air (by making use of the large pressure difference at the start of injection) and afterwards the argon flow is set to the nominal one (which marks the time $t= 0$~h in simulation). In the absence of convection, the triple-hole distributor proposed in geometry $B$ removes O$_2$ concentration very efficiently from the region between distributors, and then the fresh argon gently displaces the Ar/O$_2$ admixture upwards. The relatively low diffusion of O$_2$ at high pressure and its lower mass favours its extraction from the uppermost outlets.

\begin{figure} [h!!]
    \centering
    \begin{minipage}{0.6\textwidth}
       \centering
        \includegraphics[width=1\linewidth]{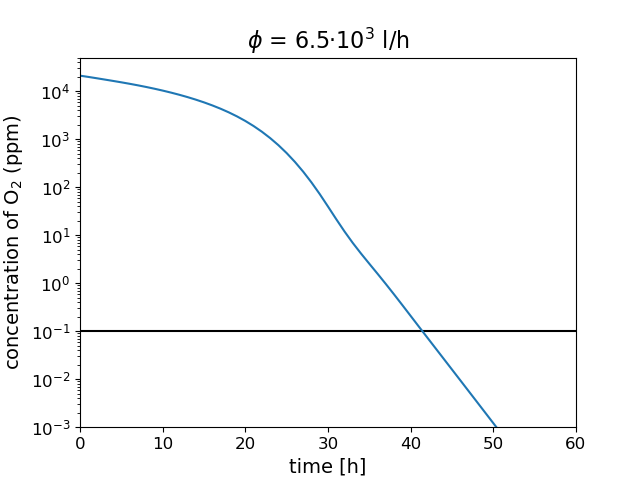}
    \end{minipage}\hfill
    \caption{Evolution of the oxygen concentration upon starting the gas flow in the TPC ($\Phi = 6.5$ m$^3$/h, $P=10$~bar), for distributor geometry $B$. The initial value corresponds to a 21\% O$_2$ concentration {in air at 1~bar, in a mixture of air at 1~bar with 9~bar of argon}. The line shows the 0.1~ppm landmark, corresponding to a 10\% charge loss in a cathode-anode electron transit.}\label{O2_conc}
\end{figure}

Fig.~\ref{O2_conc} shows the concentration of oxygen integrated over the chamber as a function of time. It is worth noting that, in the limit of large diffusion (where O$_2$ would be uniformly distributed over the chamber at all times), the time constant for O$_2$ replacement would be directly $\tau_{_{\text{O}_2}} = V/\Phi$ (see appendix, {section~\ref{filling}}), with $V$ being the chamber volume (100~m$^3$). Reaching the 0.1~ppm landmark value, for instance, would take 188~h, approximately 4.5 times longer than observed in the CFD simulations (42~h). 

\begin{figure}[h!!!]
    \begin{minipage}{0.5\textwidth}
        \centering
        \includegraphics[width=2\linewidth]{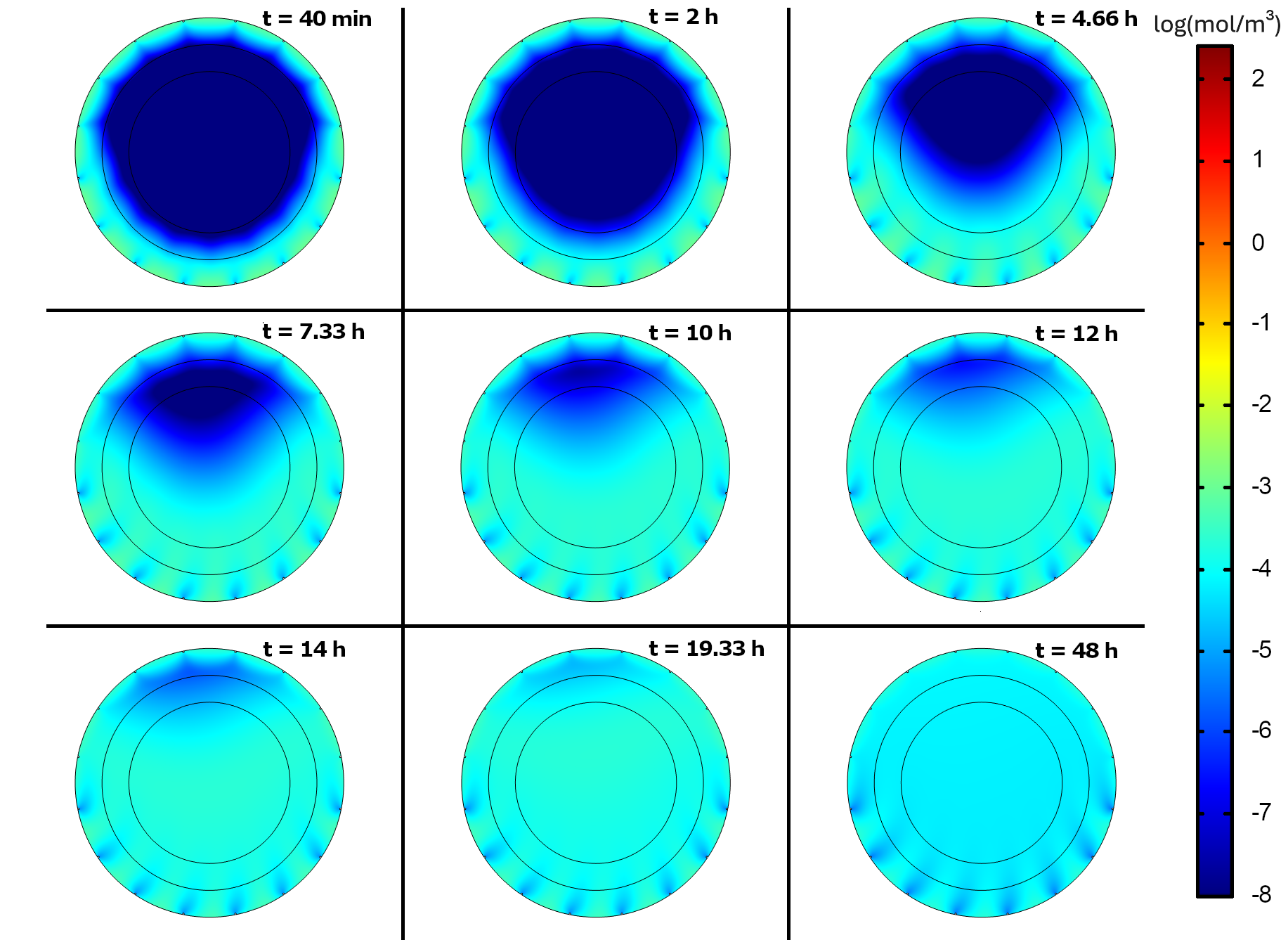}
    \end{minipage}
    \hfill
    \caption{Time lapse of the distribution of O$_2$ concentration (logarithm of the concentration in mol/m$^3$) in our reference 10~bar Ar/CF$_4$ TPC for geometry $B$ (holes in `triplet' configuration), under typical outgassing and flow rate conditions. The concentric circles mark the limits of the different mesh regions. The transient phase of the gas mixing can be observed mostly during the first 20~h. After that, the system approaches steady state and the concentration becomes homogeneous in the chamber. The images show clearly how the gas mixing is driven in these conditions by the velocity field, going from bottom to top.} \label{2Dconc}
\end{figure}

\subsection{Oxygen contamination}

When studying the evolution of oxygen, two outgassing rates have been considered: i) $Q=10^{-8}$ $\frac{\textnormal{mbar}\cdot{\textnormal{l}}}{\textnormal{cm}^2\cdot{\textnormal{s}}}$ and ii) $Q=0.05\times10^{-8}$ $\frac{\textnormal{mbar}\cdot{\textnormal{l}}}{\textnormal{cm}^2\cdot{\textnormal{s}}}$. Similarly, for the gas flow: i) $\Phi=6.5$ m$^3$/h and ii) $\Phi=20$ m$^3$/h. 
 The outgassing rate was assumed constant and equal to its value at 10~h. This has a dual benefit: i) it provides a safe upper limit to the O$_2$ concentration (compared to the anticipated $t^{-1/2}$ law) and ii) it allows a straightforward interpretation of when the system achieves stationary conditions. Details are compiled in Table~\ref{OutTab}.

\begin{figure}[h!!!]
    \begin{minipage}{0.5\textwidth}
        \centering
        \includegraphics[width=2\linewidth]{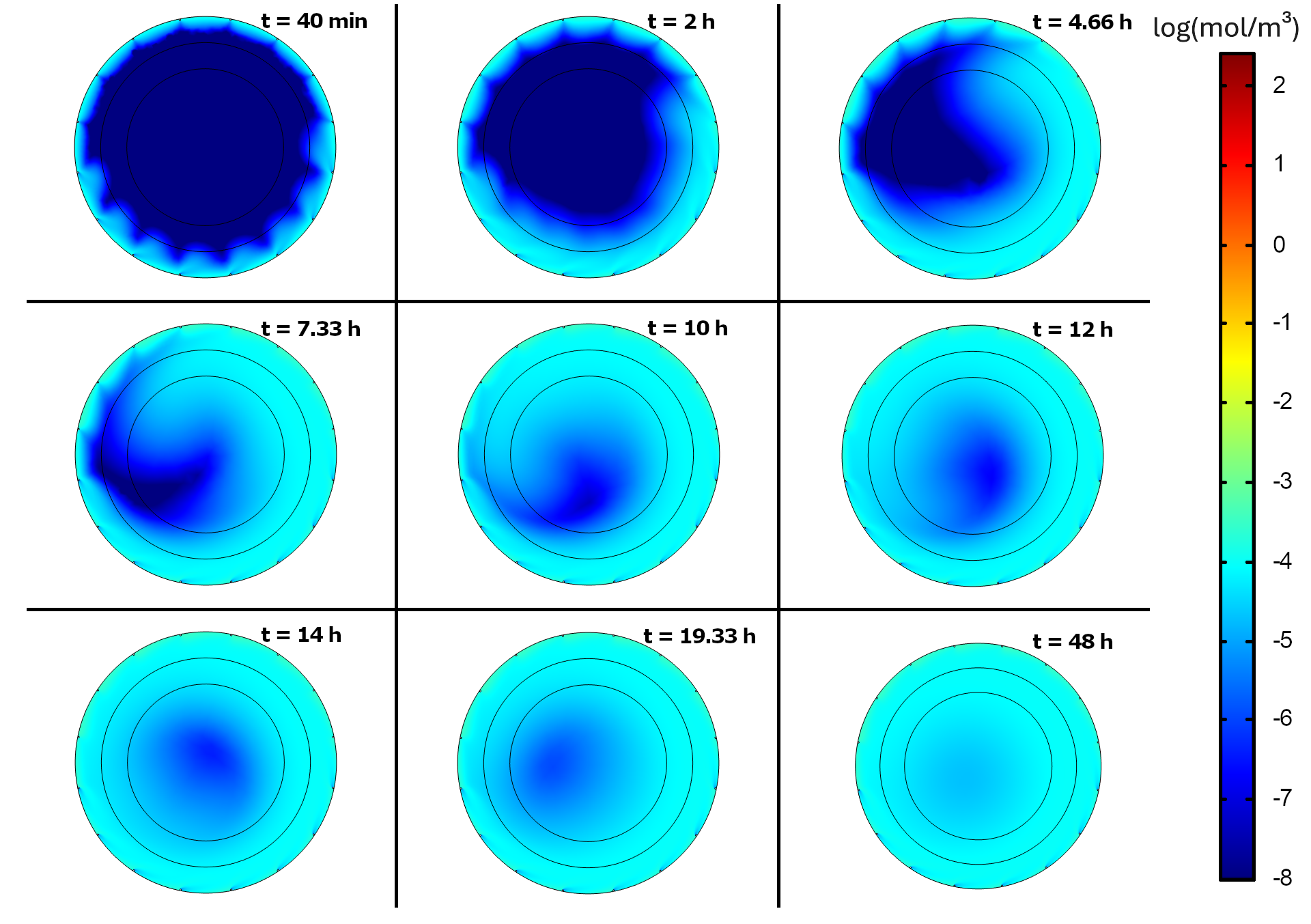}
    \end{minipage}
    \hfill
    \caption{Time lapse of the distribution of O$_2$ concentration (logarithm of the concentration in mol/m$^3$) in our reference 10~bar Ar/CF$_4$ TPC for geometry $A$ (radial holes), under typical outgassing and flow rate conditions. After 24~h, steady state conditions have not been achieved yet.} \label{2Dvelo_1h}
\end{figure}

\begin{figure} [h!!]
    \centering
    \begin{minipage}{0.5\textwidth}
       \centering
        \includegraphics[width=1\linewidth]{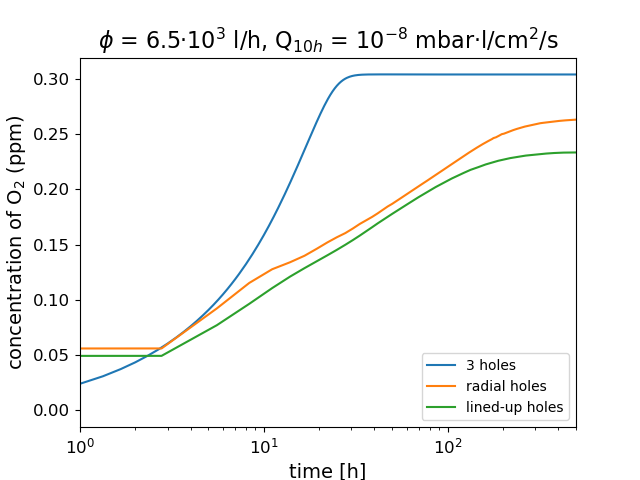}
    \end{minipage}\hfill
    \begin{minipage}{0.5\textwidth}
        \centering
        \includegraphics[width=1\linewidth]{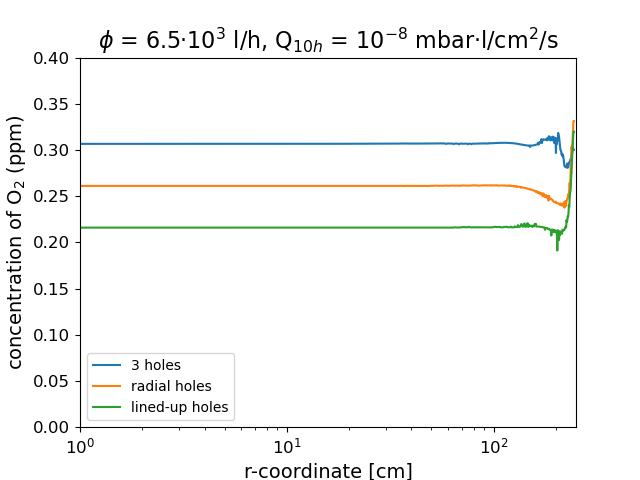}
    \end{minipage}
    \caption{Oxygen concentration simulated up to 500~h in our reference 10~bar Ar/CF$_4$ TPC, under typical outgassing and flow rates, for three different distributor geometries (three holes: blue; single radial holes: orange; single lined-up holes: green). Left: evolution of the total O$_2$ concentration with time. Right: radial distribution (logarithmic x-scale) of O$_2$ at the 500~h mark, with all distributor geometries displaying a mostly homogeneous contaminant distribution. (The contribution from the outermost 5~cm of the chamber has been removed.)}
    \label{outgas_500h}
\end{figure}

Fig.~\ref{2Dconc} shows the evolution of the O$_2$ concentration under typical outgassing and flow rate conditions for geometry $B$ (triple-hole distributor). For comparison, Fig.~\ref{2Dvelo_1h} shows the results for geometry $A$ (single-hole distributor) when the holes are radially-oriented. This latter case displays the formation of a vortex. Therefore, while in the default geometry $B$ (hole triplet) the contaminants are uniformly mixed quickly thanks to the velocity field, in geometry $A$ (hole singlet) the uniformity is achieved through diffusion, involving considerably longer times. 
Single or double vortices are prevalent in single-hole distributor geometries, for the conditions studied, but are absent in the triple-hole ones. The implications of this observation can be summarised in Fig.~\ref{outgas_500h}. In the left panel, the time evolution of the O$_2$ concentration is shown for three different distributor geometries (triple holes: blue; radial holes: orange; lined-up holes: green), corresponding to geometries $A$ and $B$ in table~\ref{DistTab}.
The right panel shows the radial distribution of impurities (logarithmic X-scale) when all geometries have achieved steady state (500~h): it is completely homogeneous except in the outermost regions, where it can vary up to 30\%. The triplet configuration reaches stationary conditions within 20~h, compared to almost 10 times that time for the 1-hole geometries.\footnote{The stationary limit following eq.~\ref{Sol_OR_ctant_2} in appendix {(section~\ref{constant_OR})} is $\frac{Q\cdot{A}}{P\cdot\Phi} = 0.43$~ppm, to be compared with the simulated one of 0.31~ppm. The deviation can be attributed to i) the uncertainty in the flow estimate from the average velocity at the inlets and ii) the contribution from the last 5~cm of the chamber, that has a complex dynamics and has been removed here aiming at a simplified discussion (the last cm's from the chamber are usually excluded from any physics analysis, as discussed earlier in text).}

\begin{figure} [h!!!]
    \centering
    \begin{minipage}{0.49\textwidth}
        \centering
        \includegraphics[width=1\linewidth]{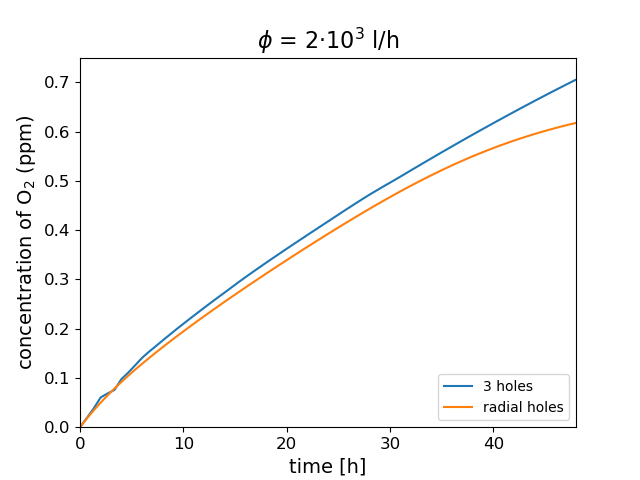}
    \end{minipage}
    \begin{minipage}{0.49\textwidth}
        \centering
        \includegraphics[width=1\linewidth]{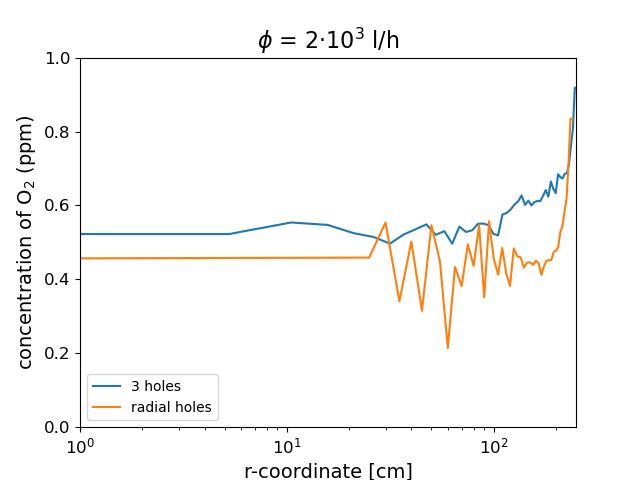}
    \end{minipage}
    \begin{minipage}{0.49\textwidth}
       \centering
        \includegraphics[width=1\linewidth]{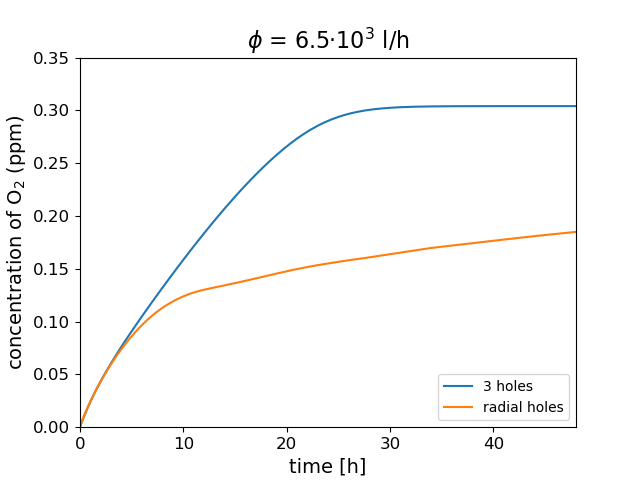}
    \end{minipage}\hfill
    \begin{minipage}{0.49\textwidth}
        \centering
        \includegraphics[width=1\linewidth]{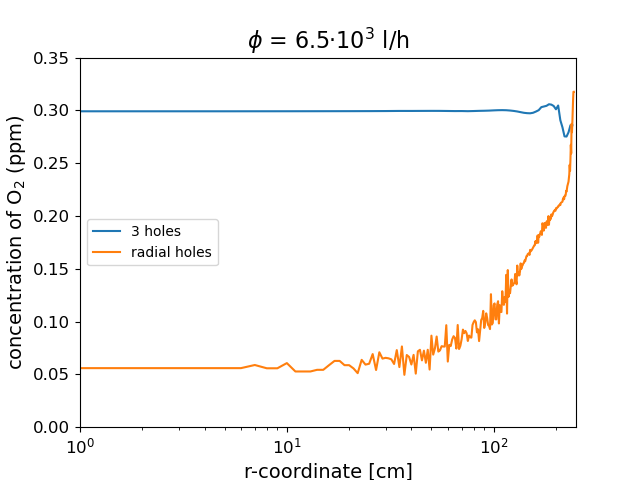}
    \end{minipage}
    \begin{minipage}{0.49\textwidth}
       \centering
        \includegraphics[width=1\linewidth]{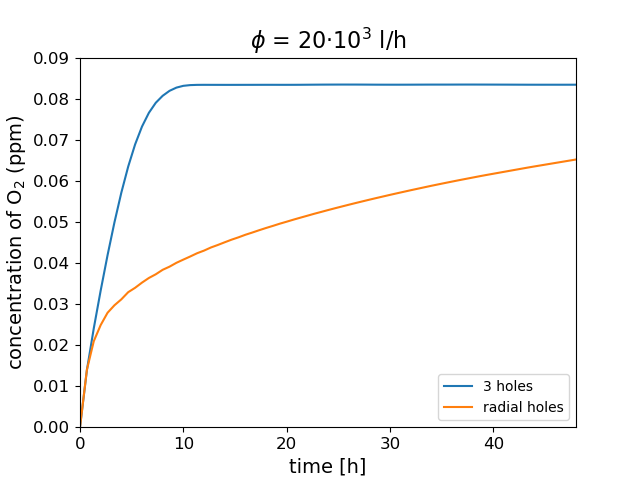}
    \end{minipage}\hfill
    \begin{minipage}{0.49\textwidth}
        \centering
        \includegraphics[width=1\linewidth]{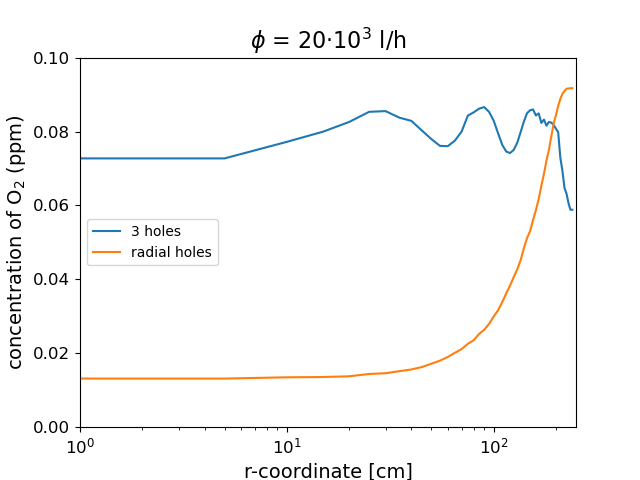}
    \end{minipage}
    \caption{Time evolution of the O$_2$ contamination (left), and radial profile at the 48~h mark, in our reference 10~bar Ar/CF$_4$ TPC. Two distributor geometries are considered (triple-holes in blue, radial-holes in orange). From top to bottom, the gas flow is 2~m$^3$/h, 6.5~m$^3$/h and 20~m$^3$/h, for a typical value of the outgassing rate. Steady state conditions for the triple-hole case are reached at about 70~h (top), 25~h (middle), and 8~h (bottom). Right: radial distributions reflecting, in the single-hole case, how impurities reach into the central region of the TPC mostly through diffusion from large radii. The wiggling radial pattern observed at high flow for the triple-hole case mirrors the position of the gas inlets. (The contribution from the outermost 5~cm of the chamber has been removed.)}
    \label{High_Ar_flow}
\end{figure}

It is difficult to assess the significance of the greater tendency to vortex formation in the single-hole distributors, and whether it will be maintained in an actual experiment when considering the exact geometry and the unavoidable convection currents stemming from temperature gradients. Notwithstanding, the triple-hole distributor is vortex-free and consistently achieves faster mixing with contaminants as well as more efficient sweeping of the outgassing from the chamber walls. This is illustrated in Fig.~\ref{High_Ar_flow}, where three different gas flows are compared under typical values of the outgassing rate (top figure: 2 m$^3$/h, middle figure: 6.5 m$^3$/h, bottom figure: 20 m$^3$/h). The time evolution is displayed, up to 48~h, on the left column, while the radial distribution of contaminants is shown on the right column. Due to the low gas flow in the top figure (and correspondingly low discharge velocity), the time scale is too short to lead to vortex formation and the two distribution schemes perform similarly up to nearly the steady state value (1.4~ppm for these conditions). Even in the absence of vortices, the triple-hole distribution is more efficient at mixing due to better sweeping of the chamber walls. For larger gas flows (middle and bottom figures), the fluid dynamics of the radial-hole geometry is dominated by the formation of a central vortex, with impurities reaching the central region through diffusion alone (right figures).

\begin{figure} [h!!!]
    \centering
    \begin{minipage}{0.8\textwidth}
       \centering
        \includegraphics[width=1\linewidth]{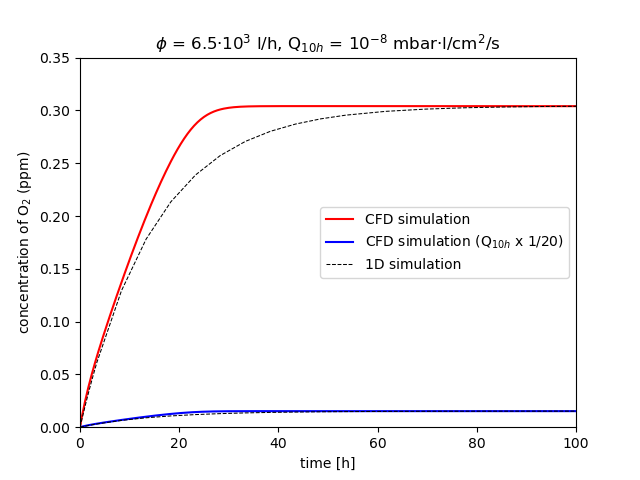}
    \end{minipage}
    \caption{Time evolution of the O$_2$ contamination in our reference 10~bar Ar/CF$_4$ TPC for `typical' outgassing and flow rates, in the default geometry $B$ (holes in `triplet' configuration) in red. A 1-D solution (`infinite diffusion' limit) is shown in dashed lines. The CFD solution for `optimistic' outgassing values (blue) reflects a simple down-scaling, without altering the shape. {See table \ref{OutTab} for typical and optimistic values.} (To account for the numerical deviation from the steady-state limit, as discuss earlier in text, the outgassing in the 1-D solution has been downscaled by 28\%.)}
    \label{1D_ctant}
\end{figure}

The results obtained in this section can be analysed by comparison with the one-dimensional solutions, which effectively consider an `infinite diffusion' scenario (see appendix, {section~\ref{constant_OR}}). Fig.~\ref{1D_ctant} shows how the observed time-dependence (red line) is reasonably well approximated by the 1-D solution (dashed line), implying that the achieved mixing rate is compatible with the ideal limit of perfect mixing. The blue line, obtained at 20 times less outgassing (`optimistic' case) aims at illustrating that the curves obtained by CFD reflect a global scaling factor for different outgassing rates, following as well the expectation from the 1-D solutions. 
The asymptotic value in Fig.~\ref{1D_ctant} (red line) of 0.3~ppm would correspond to a 30\% charge loss following the estimates in section~\ref{ImpuritiesInTPC}. Considering the range of outgassing rates in table~\ref{OutTab}, an upper limit to the charge loss after 24~h can be situated in the range 1.5-30\%, for the gas flow conditions in Fig.~\ref{ImpuritiesInTPC}.

\subsection{Water contamination}

For Teflon, the outgassing rate of water is significantly higher than that of O$_2$, so that if outgassing did not decrease over time then the necessary purity levels might never be achieved. Instead of assuming a constant value, we consider in this case the actual outgassing law based on eq.~\ref{outgasLaw}, with outgassing rates after 10~h taken from table~\ref{OutTab}: $3.5 \times 10^{-8}$ $\frac{\textnormal{mbar}\cdot{\textnormal{l}}}{\textnormal{cm}^2\cdot{\textnormal{s}}}$ (`optimistic') and $10 \times 10^{-8}$ $\frac{\textnormal{mbar}\cdot{\textnormal{l}}}{\textnormal{cm}^2\cdot{\textnormal{s}}}$ (`typical').

\begin{figure} [h!!!]
    \centering
        \includegraphics[width=0.7\linewidth]{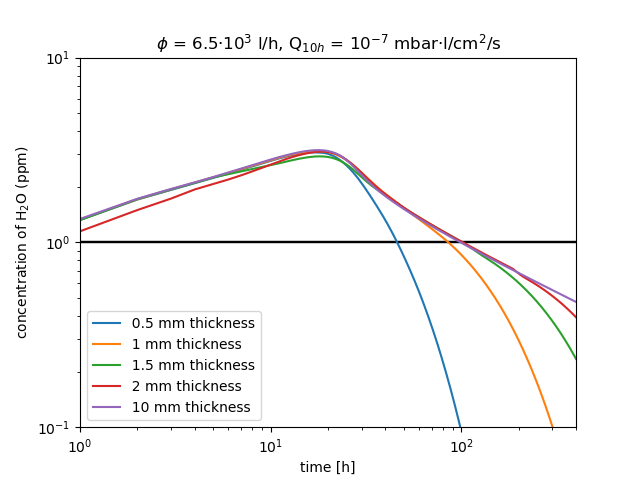}
    \caption{Evolution of water concentration as a function of time for `typical' values of the outgassing and flow rates, in our reference 10~bar Ar/CF$_4$ TPC, for geometry $B$ (triple-hole distributor). Three different Teflon thicknesses are considered. The land-mark value of 1~ppm (less than 1~mm track distortion on a 5~m drift) is achieved at 99~h in the worse-case scenario.}\label{Teflon_thick}
\end{figure}

We concentrate on the case of typical outgassing and flow rates, and five different Teflon thicknesses, including the case where Teflon would be so thick as to preserve a $t^{-1/2}$ behaviour throughout. Based on the analysis in the previous section, the configuration with triple-holes has been chosen. Results are shown in Fig.~\ref{Teflon_thick}. For thicknesses below 10~mm, a transition between the $t^{-1/2}$ and exponential regimes can be observed. As the outgassing rates of H$_2$O are about 10 times larger than those of O$_2$ in the case of Teflon, the time scales involved are larger too. The horizontal line shows the 1~ppm-level, which is a plausible requirement to keep space resolution undistorted below 1~mm, based on the analysis presented in section~\ref{ImpuritiesInTPC}. For a thin-Teflon reflector down to 0.5~mm thickness, coupled to an ESR reflective coating (see e.g.~\cite{JUSTO-REF}), that landmark value is achieved in 46~h, while it reaches up to 99~h as the Teflon becomes thicker. The result is sobering: when considering the displacement of air during filling and the time needed for water (oxygen) to fall below the concentration that makes its contribution negligible for tracking (attachment), about 6-10 days would be typically needed. Well-defined cleaning protocols and surface treatment, as discussed for example in \cite{Poole}, together with an increase of the gas flow above the levels discussed here, or designing the system in compliance with the requirements needed for pumping, could plausibly reduce these times to the scale of about 1-2 days. Nevertheless, additional (uncontrolled) outgassing sources would render the operation of the detector problematic.

\begin{figure} [h!!]
    \centering
    \begin{minipage}{0.5\textwidth}
       \centering
        \includegraphics[width=1\linewidth]{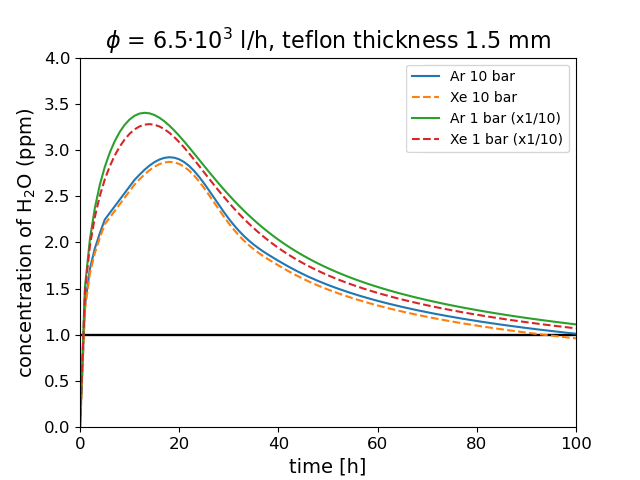}
    \end{minipage}\hfill
        \begin{minipage}{0.5\textwidth}
       \centering
        \includegraphics[width=1\linewidth]{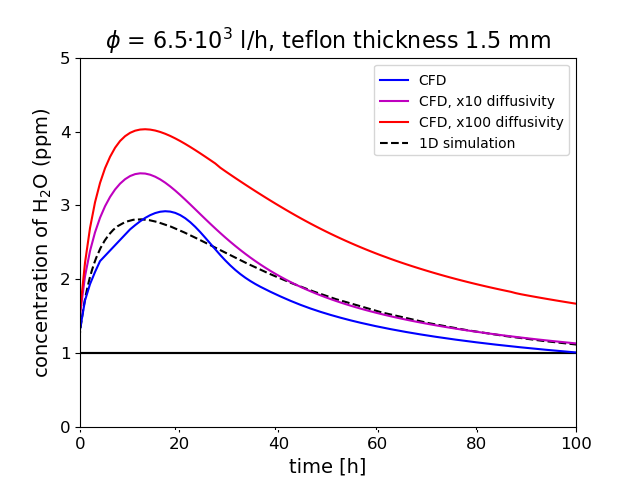}
    \end{minipage}\hfill
    \caption{Left: water concentration as a function of time for `typical' values of the outgassing and flow rates, in our reference 10~bar Ar/CF$_4$ TPC, assuming a Teflon thickness of 1.5~mm. Different gas species (Ar: continuous; Xe: dashed) and pressures have been considered. Right: time-dependence for Ar/CF$_4$ at 10~bar (blue) and comparison with results assuming an artificially increased diffusion coefficient of $\times 10$ (purple) and $\times 100$ (red). Dashed-lines correspond to a 1-D model (`infinite diffusion' limit).}
    \label{WaterSystematics}
\end{figure}

Additional studies are presented in Fig.~\ref{WaterSystematics} in order to illustrate the main dependencies. On the left figure, the behaviour for different gases is shown for a fixed flow rate of 6.5 m$^3$/h (argon: continuous lines; xenon: dashed lines). It can be seen how, despite the diffusion coefficient of H$_2$O being larger in argon by about 50\% (table~\ref{OutTab}), the impact on the time-evolution of the impurities is almost indiscernible. When reducing the pressure down to 1~bar, the 10-fold increase in the diffusion coefficient accelerates the mixing as expected; however, the reduction in the position of the maximum is a mere 30\%. As the number of normal litres is accordingly reduced by a factor of 10 in those conditions, the concentration of impurities increases in inverse proportion (the Ar and Xe lines at 1~bar have been downscaled by a factor of 10).

Fig.~\ref{WaterSystematics}-right shows a comparison with the 1-D solutions for the case of time-dependent outgassing (see appendix, {section~\ref{powerLaw_OR}}). As for the case of constant outgassing, the results compare well with the 1-D model, although the agreement is qualitatively worse. Artificially increasing the gas diffusion accelerates the mixing, but very mildly in the conditions studied (as indirectly shown in Fig.~\ref{WaterSystematics}-left). The conclusion is once again that the velocity field in combination with gas diffusion can distribute the outgassed species uniformly over the chamber for the optimised `triplet' distributor, as seen in Fig.~\ref{2Dconc}. Also in this case, simulations for lower (`optimistic') outgassing rates display an exact down-scaling compared to the solutions presented in Fig.~\ref{WaterSystematics}, as expected from the 1-D solutions.

\subsection{Nitrogen contamination}

Nitrogen can be removed from the system by flushing the main gas through `hot' (temperature-activated) purifiers/getters, a common asset for the operation of noble gas detectors. The process of nitrogen purification becomes challenging, however, in case of operation under large gas flows (that could modify the getter temperature) and/or in the presence of molecular additives such as CF$_4$ (that may react with the getter materials). Following conventional wisdom, and as long as a purification system compatible with Ar/CF$_4$ operation is not demonstrated at the necessary scale, we shall conservatively assume that N$_2$ accumulates in the TPC as time passes.

Fortunately, the tolerance to N$_2$ is by far the largest of common outgassing species, up to 10~ppm (table \ref{OutTab}). Based on the time profile of the outgassing rate (eq. \ref{outgasLaw}), the range of outgassing values in table \ref{OutTab}, and the chamber volume, the evolution of N$_2$ inside the TPC can be approximated as (see appendix, {section~\ref{accumulation}}):
\begin{equation}
    f_{\text{N}_2}\sim(3-9)\textnormal{ppm} \cdot \sqrt{t[\textnormal{days}]}, \label{N2acc}
\end{equation}
with the pre-factor covering the range dubbed earlier as `optimistic' (3~ppm) and `typical' (9~ppm). In practice, this expression indicates that N$_2$ can approach harmful concentrations quite quickly, within one day; however, the concentration will increase slowly thereafter. Assuming 10~years of continuous operation, N$_2$ concentration would approach 180-540~ppm by the end of the experiment. In the absence of purification, the equivalent of 18-54 ND-GAr volumes of fresh gas would need to be added during the detector lifetime, to keep the N$_2$ concentration within bounds. Such an estimate represents an upper limit since: i) the outgassing rate of plastics is expected to fall exponentially as the material becomes N$_2$-depleted (eq. \ref{outgasLaw}), and ii) the impact of N$_2$ contamination on the tracking capabilities might be correctable with adequate calibration systems. Also, as discussed earlier, these concentrations are unlikely to create problems for the optical signal.

There are several ways to address the most convenient commissioning strategy, for which a reliable experimental estimate of the N$_2$ outgassing rates would be essential, first and foremost. As an example, the system could be operated in pure argon gas with both hot and cold getters for about 10~days, prior to CF$_4$ injection. N$_2$ concentration would then reach back to 10~ppm in another 10~days for `typical' outgassing values, and 30~days for `optimistic' estimates (eq. \ref{N2acc}). Injection of fresh gas in the range of 0.3-1\% after the first 10~days would guarantee operation around or below the 10~ppm N$_2$ benchmark. In the presence of calibration systems which would allow for the correction of track distortions as large as 1~cm (100~ppm of N$_2$), few additional ND-GAr volumes (in the range of 2-6) would be needed.

As CF$_4$ has a very high GWP, it seems unavoidable that `used' gas needs to be cryorecovered and CF$_4$ distilled prior to disposal to the atmosphere, whilst fresh gas is being injected.

\section{Summary and conclusions} \label{conclusion}

In this work we have addressed the problem of gas distribution and impurity mitigation for next-generation gas Time Projection Chambers in the fields of Neutrino Physics and Rare Event Searches. Specifically, we focused on the performance of an Ar/CF$_4$ (99/1) optical TPC of 5~m-length/5~m-diameter at 10~bar, with a field cage covered with Teflon lining, as recently suggested in \cite{papersOnArCF4}. We have considered an axial gas distribution scheme (along the electric field direction), that is more independent of the specific end-cap design as well as more easily tractable through computational fluid dynamics (CFD) simulations.

Distributor rods are modelled through a combined approach based on CFD simulations and generalised continuity equations, which allow a straightforward optimisation. The approach demonstrates that, for the flow rates of relevance to next-generation TPCs, a nearly-perpendicular discharge angle (less than 5$^\circ$ with respect to the normal) can be achieved for hole diameters at the mm-diameter scale, thus ensuring a priori a good bottom-up circulation of the gas for any position along the axial direction, without the creation of gas pockets. Holes are separated by 10-50~mm, the rod diameters being in the range of 30-70~mm (possibilities for further reducing this diameter are given in the text, although they are outside the scope of this work). The discharge velocity of the holes and the pressure drop at the distributor rod are generally far from the regime where they could pose problems, and are not critical in the optimisation.

The proposed gas distribution, based on eight perforated plastic rods (capped at their ends) for injection and eight for extraction, can efficiently sweep the outgassing from the walls when a triple-hole perforation is implemented (one hole situated along the radial direction and two azimuthally). By contrast, single-hole distributors perform worse at sweeping the gas from the walls and are more prone to vortex formation, resulting in much longer times to achieve steady-state conditions. Additionally, the triple-hole configuration leads to an efficient displacement of air during chamber filling (down to 0.1~ppm of O$_2$) in about 42~h, which is 4.5 times faster than expected from a diffusion-dominated 1-D model. (Assuming no pocket formation, the 1-D model provides an estimate of the longest extraction time).

The contamination of O$_2$ (H$_2$O) during chamber operation was studied, assuming circulation in closed loop with the gas forced into conventional `cold' getters. For a realistic flow rate of 6.5~m$^3$/h at 10~bar, a maximum impurity level of 0.3~ppm (3~ppm) would be achieved in just over 24~h for typical values of the outgassing rates ($Q_{\text{10h}} \simeq 10^{-8}\frac{\textnormal{mbar} \cdot l}{\textnormal{cm}^2 \cdot \textnormal{s}}$ for O$_2$, and $Q_{\text{10h}} \simeq 10^{-7}\frac{\textnormal{mbar} \cdot l}{\textnormal{cm}^2 \cdot \textnormal{s}}$ for H$_2$O). These concentrations translate into maximum charge losses of about 30\% (due to O$_2$) and track distortions of 3~mm (due to H$_2$O). The deterioration in the TPC response is reduced by a factor of 3 after 100~h, approximately following the anticipated reduction of the outgassing rate for technical plastics, as $\sim t^{-1/2}$. Generally speaking, and based on our experimental survey of the Teflon outgassing rates, after 100~h the expected losses due to attachment would be situated in the range 0.5\%(`optimistic')-10\%(`typical'), with track distortions ranging between 0.3~mm(`optimistic') and 1~mm(`typical'). In the proposed gas distribution scheme, for both O$_2$ and H$_2$O, mixing with the main TPC gas takes place at a pace compatible with the expectation from a diffusion-dominated 1-D model, which provides a benchmark for the shortest mixing time achievable in practice. Importantly, this close-to-ideal mixing renders a uniform distribution of contaminants within merely 24~h, a relevant feature in the need to apply corrections to reconstructed events.

Under the conservative assumption that N$_2$ purification at high flows and in the presence of CF$_4$ needs dedicated studies and cannot be taken for granted at the moment, we discussed mitigation strategies to the problem of N$_2$ accumulation. It is shown that a total of fresh gas worth 18(`optimistic')/54(`typical') TPC volumes would be needed during 10~years of continuous operation, a requirement that can be met, for instance, by injecting fresh gas at a mere 0.3-1\% fraction of the flow of the main gas line. If track distortions of up to 1~cm (100~ppm of N$_2$) could be corrected online through (laser-based or other similar) calibration systems, the additional fresh gas would be situated in the range of 2(`optimistic')-6(`typical') additional TPC volumes. These values are comparable to those needed during chamber filling, if TPC evacuation were not possible.

{ Some considerations towards realism are pertinent. First, a complete 3D simulation of the final system including the end-caps and thermal sources (to account for possible convection currents), as well as the possible thickening of the distributor walls, might suggest variations of the gas distribution proposed, e.g. considering distributor rods of smaller diameter or the possibility of placing them behind the field cage reflector (therefore abandoning the triplet configuration). Second,
t}hese conclusions, as well as most of the work in the body of the paper, focuses on `typical' values of the outgassing and gas flow rates, but there is room for improvement, and lower outgassing values might be achieved through suitable cleaning/surface finish protocols, whilst higher gas flows will likely become affordable in a few-year timescale. Ultimately this work is aimed at showing that, while operation of the optical TPC discussed here seems perfectly viable from the point of view of material compatibility, detailed outgassing assays of the materials involved in construction seem crucial to ensure the chamber performance and the necessary mitigation strategies.
\acknowledgments

This research has been funded by the Spanish Ministry (‘Proyectos de Generación de Conocimiento’, PID2021-125028OB-C21, PID2021-125028OB-C22). It has also received complementary financial support from the regional government Xunta de Galicia (Centro singular de investigación de Galicia accreditation 2019-2022), and by the national program “María de Maeztu” (Units of Excellence program MDM-2016-0692). We thank CERN for computing resources. The authors are grateful to the reviewer, whose comments helped improving the manuscript.

\appendix

\section{Iterative implementation of a multi-hole distributor} \label{iterative_script}

\begin{figure}[h!!!]
    \centering
    \includegraphics[scale=0.62]{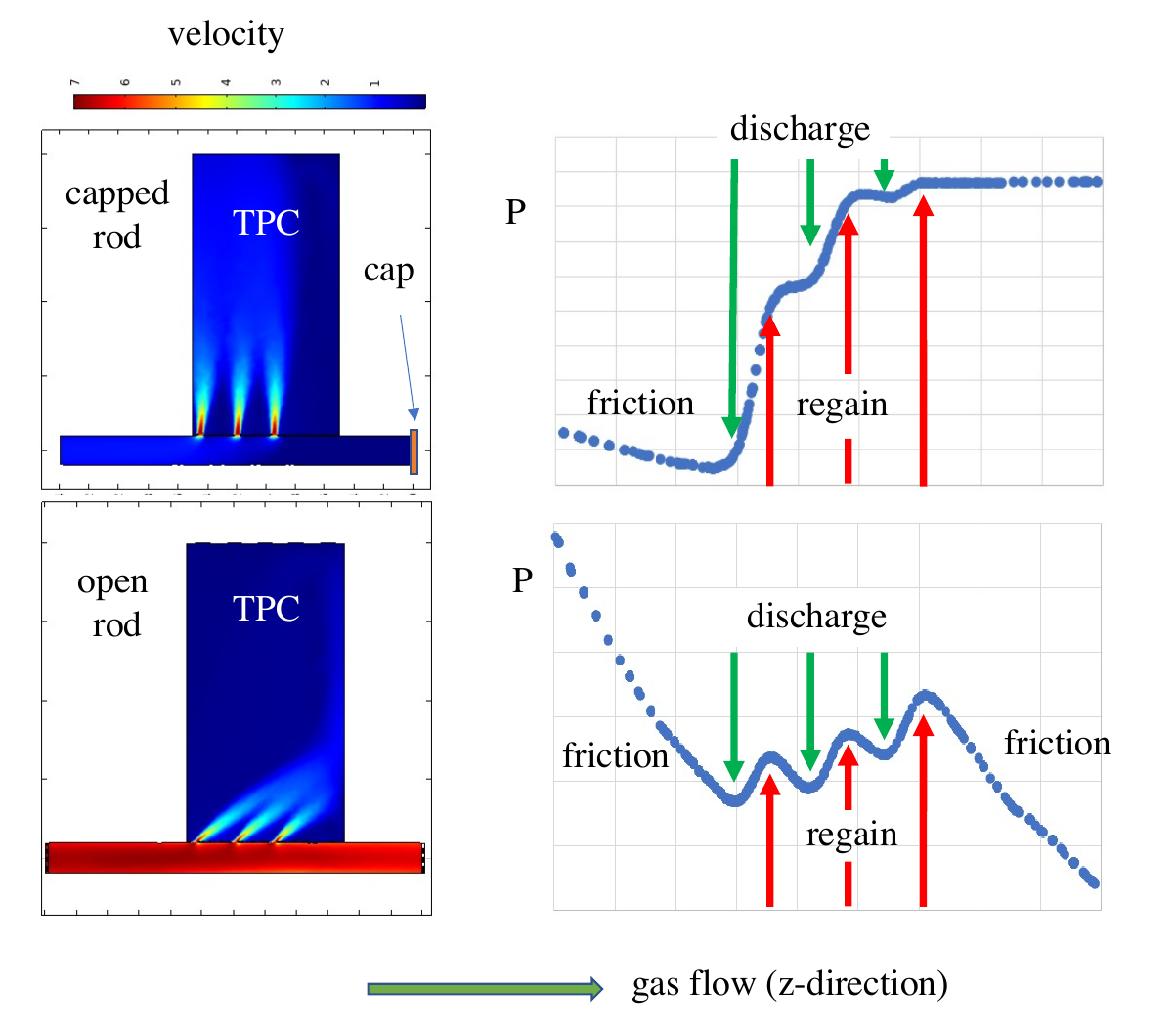}
    \caption{Left: exemplary three-dimensional CFD simulations for a 3-hole distributor (top: capped; bottom: open), representing the modulus of the gas velocity in the plane containing the holes. Right: pressure as a function of the longitudinal position, showing the discharge and regain zones. Friction dominates outside the holes' region.} 
    \label{COMSOL_1D}
\end{figure}

Results from a three-dimensional CFD simulation performed, illustratively, for a 3-hole distributor, are shown in Fig.~\ref{COMSOL_1D}. The figure provides a cross-section at the holes' middle plane (the modulus of the gas velocity is represented in the left column, the pressure profile in the right one). Two reference cases are considered: a capped distributor (top) and an open one (bottom). As expected, an open distributor is less efficient, given that only a fraction of the gas exits through the holes and into the chamber—as little as a few \% for high gas flow. 
It also leads to a higher pressure drop along the distributor due to friction, significantly greater directionality of the outgoing gas and a non-uniform discharge pattern. Low efficiency in combination with a directional and non-uniform discharge pattern will adversely affect the distribution of impurities throughout the chamber and the time required to achieve uniform mixing with the main gas. Therefore, we opted in text for a capped distributor, ensuring uniform and normal injection of all the circulated gas.

\begin{figure}[h!!!]
    \centering
    \includegraphics[scale=0.8]{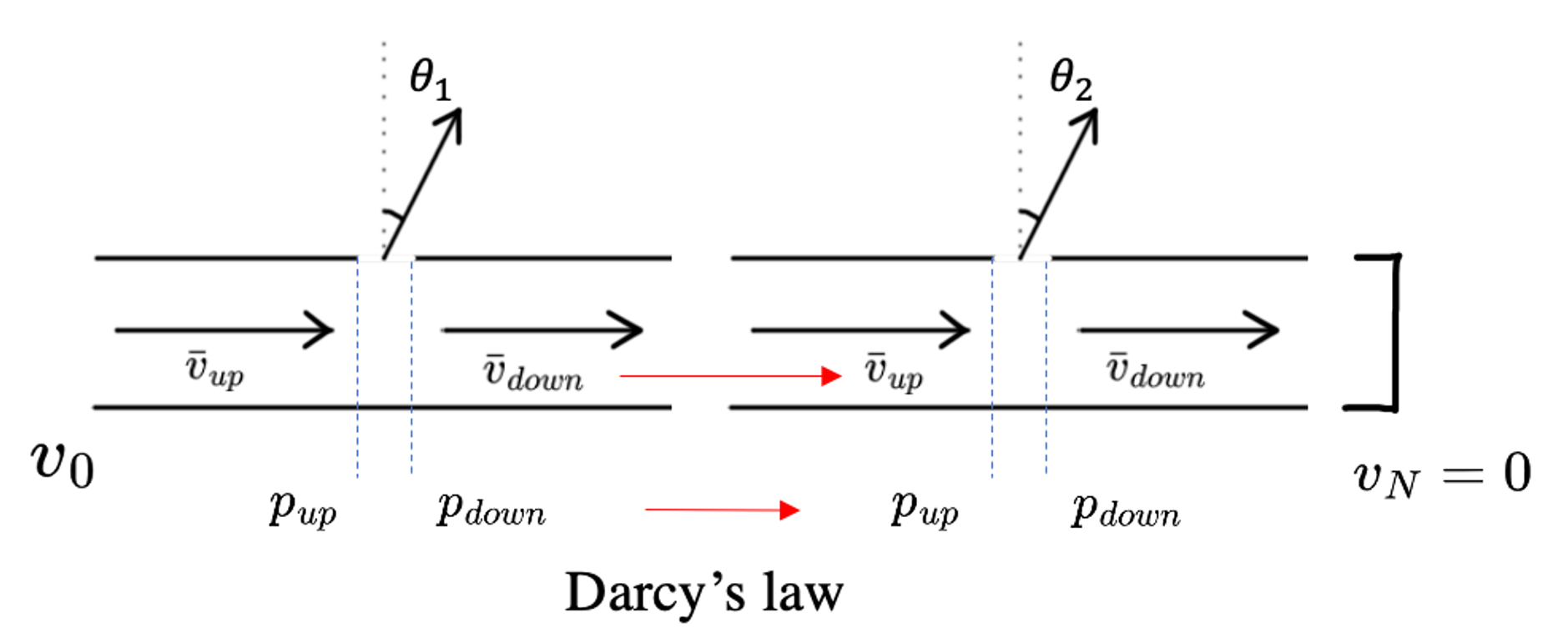}
    \caption{Perforated gas distributor capped at its end, together with the main magnitudes introduced in text. The sketch also gives an approximate representation of the solving method, that tries an initial value $P_\text{u}$ upstream of the first hole, for an assumed flow, and iterates until a null velocity is obtained at the end of the distributor.} 
    \label{Distri}
\end{figure}

The mass and momentum -balance equations for each hole section (eqs. \ref{Rec1} and \ref{Rec2}) can be solved iteratively: for an assumed gas flow (velocity) at the entrance of the distributor, the pressure upstream of the first hole is scanned till the velocity downstream of the last hole converges to zero (Fig. \ref{Distri}). The set of hole-equations that need to be iteratively propagated along the distributor is:
\begin{eqnarray}
\frac{P_i+P_{i+1}}{2}&=&P_{_\text{TPC}} + \frac{\rho D^4(v_i-v_{i+1})^2}{2d^4C_{d,i}^2} \label{A1}, \\
P_{i+1}-P_i&=&C_{r,i}\frac{1}{2}\rho(v_i^2-v_{i+1}^2). \label{A2} 
\end{eqnarray}
C$_d$, C$_r$ are dependent on the gas properties and little-dependent on geometry. We adopted the formalism in \cite{Bailey}, in which C$_d$ is parameterized as a function of the ratio $\frac{1}{2}\rho v_\text{u}^2/\Delta{P}_\text{hole}$ and C$_r$ as a function of $\Delta{v} = v_\text{u}-v_\text{d}$.

For a complete description, equations \ref{A1}-\ref{A2} must be coupled hole-by-hole after accounting for any loss due to friction. The pressure drop in between hole sections is included through Darcy's friction law:
\begin{equation}
P_{\text{u}(i+1)} - P_{\text{d}(i)} = f_D\frac{\Delta{L}}{D}\rho\frac{{v_{\text{d}(i)}}^2}{2}, \label{fricc}
\end{equation}
where $\Delta{L}$ is the distance between holes, $\rho$ the gas density and $f_D$ is Darcy's friction coefficient. A distinction is made, as customary, between laminar ($Re<4000$) and turbulent ($Re>4000$) regime, with $Re$ being the Reynolds number:
\begin{equation}
Re = v_{\text{d}(i)} \cdot \Delta{L} \cdot \frac{\rho}{\mu},
\end{equation}
and $\mu$ is the (dynamic) viscosity of the fluid. For laminar flow $f_D=64/Re$ is assumed and, for turbulent flow:
$$\frac{1}{\sqrt{f_D}}=-2\log\left(  \frac{k_s}{3.7D}+\frac{2.51}{Re\cdot\sqrt{f_D}}\right).$$
Here $k_s$ is the roughness of the pipe wall (a typical value for a generic acrylic material was assumed: $k_s=1.1$~$\mu$m \cite{roughness}).

\section{Useful analytic limits for capped distributors} \label{analytic}

The solution to $v_d$, for an initial upstream velocity $v_u$, can be easily obtained from eqs. \ref{Rec1} and \ref{Rec2}:

\begin{equation}
v_d = \frac{\frac{D^4}{d^4}\frac{1}{C_d^2}\rho{v_u} - \sqrt{\frac{1}{4}C_r^2\rho^2v_u^2 + 4\left( \frac{1}{2} \frac{D^4}{d^4}\frac{1}{C_d^2}\rho + \frac{1}{4}C_r\rho \right)(P_u - P_{_\text{TPC}})}}{2\left(\frac{1}{2}\frac{D^4}{d^4}\frac{1}{C_d^2}\rho + \frac{1}{4}C_r\rho \right)}.\label{vd_ana}
\end{equation}
Since $C_r$ is bound to $[0,2]$ and $C_d$ to $[0,1]$ it follows that:
\begin{equation}
\frac{1}{4}C_r << \frac{1}{2}\frac{D^4}{d^4}\frac{1}{C_d^2}\label{Cr_Cd_condition},
\end{equation}
as long as $d<<D$, which is fulfilled in the distributor proposed in text by at least one order of magnitude. This allows neglecting the right-term in both the numerator and denominator of eq. \ref{vd_ana}. The left-term inside the square root is very small, part through the above relation, part because:
\begin{equation}
\frac{1}{2} \rho v_u^2 << P_u - P_{_\text{TPC}}, \label{headRatio}
\end{equation}
which is typical of a capped distributor.\footnote{The inequalities given by eq. \ref{Cr_Cd_condition} and eq. \ref{headRatio}, with the $D/d$ ratios employed in text, cause the positive solution in eq. \ref{vd_ana} to yield $v_d>v_u$, that is nonphysical.} It has been verified during simulations that the pressure difference always exceeds the velocity term by at least a factor of $10$, and generally much more. Finally, friction losses from eq. \ref{fricc} are very small due to the smallness of the ratio $\Delta{L}/{D}$ in the cases discussed (generally, 10\% at most), so we may disregard the effect too.

It therefore follows, from the above considerations, that:
\begin{equation}
P_u - P_{_\text{TPC}} \simeq \Delta{P}_\text{hole} \simeq \frac{1}{2}\frac{D^4}{d^4}\frac{1}{C_d^2}\rho(v_d - v_u)^2,
\end{equation}
irrespective from the hole position. If we assume that the discharge is uniform for each hole, that is observed in present conditions to better than 5\%, then it also follows from mass balance that:
\begin{equation}
(v_u - v_d)\cdot\pi\left(\frac{D}{2}\right)^2 = \Phi_\text{hole} = v_\text{hole}\cdot\pi\left(\frac{d}{2}\right)^2.
\end{equation}
Under the assumption of uniform discharge, the flow per hole relates to the total flow as $\Phi_\text{hole} = \Phi/N_\text{holes}$.
The analytic expressions given in eq. \ref{ana_1}, \ref{ana_2} can be finally obtained:

\begin{eqnarray}
    \Delta{P}_\text{hole} & \simeq & \frac{1}{2} \frac{1}{C_d^2} \rho v_\text{hole}^2, \\    
    v_\text{hole} & \simeq & \frac{\Phi}{\textnormal{total hole area (t.h.a.)}}.
\end{eqnarray}
Although $C_d$ is fluid-dependent, in the pressure-dominated regime defined by eq. \ref{headRatio} it approaches the asymptotic limit discussed in text: $C_d \simeq 0.63$. 

The discharge angle can be finally evaluated from expression:

\begin{equation}
    \theta = \gamma \cdot \arctan\left(\sqrt{\frac{\frac{1}{2}\rho v_{u}^2}{\Delta{P}_\text{hole}}}\right).
\end{equation}
The maximum discharge angle corresponds to the first hole upstream, for which:
\begin{equation}
    v_{u} = \frac{\Phi}{N_\text{pipes}} \frac{1}{\pi (D/2)^2},
\end{equation}
and substituting the expression obtained previously for $\Delta{P}_\text{hole}$ in the limit of small angles, we arrive at expression \ref{ana_last} in text:
\begin{equation}
    \theta_\text{max} \simeq \gamma C_d \frac{d^2}{D^2} \frac{L_\text{dist}}{\Delta{L}}.
\end{equation}
As $\gamma$ is weakly dependent on the fluid properties and geometry (e.g. \cite{Bailey}), a value of $\gamma = 0.71$ has been assumed.

The analytical expressions presented here for $\Delta{P}_\text{hole}$, $v_\text{hole}$ and $\theta_\text{max}$ have been found to agree with the exact numerical treatment to better than a few \% for all the conditions discussed in this work. This has been illustrated for instance in Fig. \ref{1Dscript}.

\section{Time-dependent differential equations}
\label{diffeq}





\begin{figure}[h!!!]
    \centering
    \includegraphics[width=0.8\textwidth]{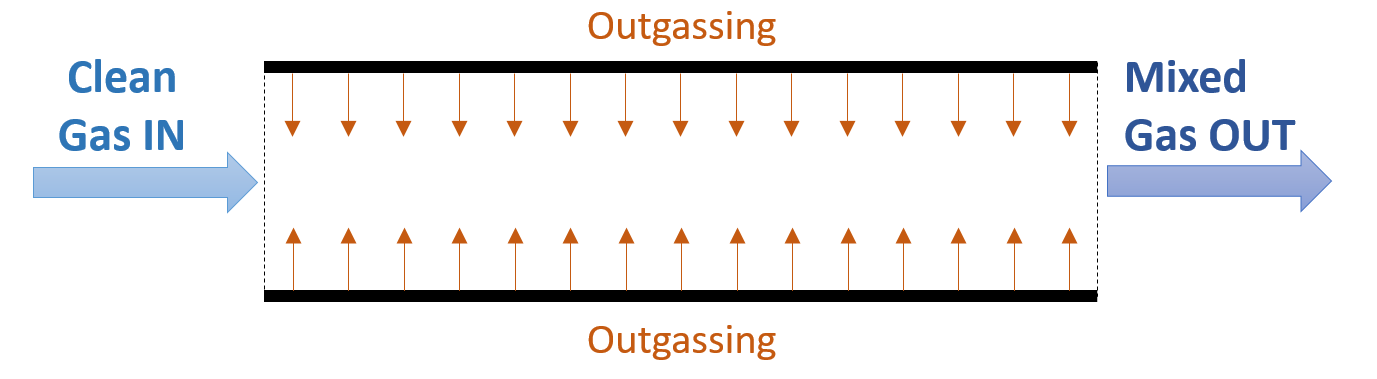}
    \caption{Schematic representation of the 1-D model used in text.} 
    \label{fig:1D_scheme}
\end{figure}

In an idealized situation, we can model the outgassing and cleaning in the control volume as a simple mass balance, as in the scheme shown in Fig.~\ref{fig:1D_scheme}.
In this case, the flow of contaminants (X) can be represented as 
\begin{equation} \label{eq:mass_flow}
    \frac{d N_\text{X}}{d t} = q(t) - \frac{d N_\text{Ar}}{d t} \cdot  \frac{N_\text{X}}{N_\text{Ar}}, 
\end{equation} 
%
%
%
where $q(t)$ represents the supply of contaminants (outgassing rate, in this case in units of number of molecules per unit time), $\frac{d N_\text{Ar}}{d t}$ is the flow rate of Ar (in this case in units of number of atoms per unit time), 
and $\frac{N_\text{X}}{N_\text{Ar}}$ is the probability that a molecule X exits the control volume (is `cleaned') at a given time. This assumption is referred in text as the `diffusion-dominated limit', under which new molecules of type X are distributed over the control volume instantly.

\subsection{Constant outgassing rate} 
\label{constant_OR}

To solve the previous differential equation, we first consider the case where the outgassing rate $q(t)$ is constant over time. Under this assumption, we can easily integrate equation \eqref{eq:mass_flow} obtaining
\begin{equation}
    N_\text{X}(t) =\frac{q}{\frac{1}{N_\text{Ar}}\frac{d N_\text{Ar}}{dt}} \, \left( 1-\exp \left\{-\frac{1}{N_\text{Ar}}\,\frac{d N_\text{Ar}}{dt} \cdot t\right\}  \right), \label{Sol_OR_ctant}
\end{equation}
once we consider that the concentration of contaminants is 0 at $t=0$.

If the contaminant is homogeneously distributed within the domain, the probability of it being cleaned by the flow of argon is proportional to its concentration. The concentration of X is then:
\begin{equation}
    f_\text{X}(t) =\frac{q}{\frac{d N_\text{Ar}}{dt}} \, \left( 1-\exp \left\{-\frac{1}{N_\text{Ar}}\,\frac{d N_\text{Ar}}{dt} \cdot t\right\}  \right). \label{Sol_OR_ctant_2}
\end{equation}
Expression \ref{Sol_OR_ctant} can be rewritten as a function of the magnitudes used in text if we recall the ideal gas law $P=\frac{N_\text{Ar}}{V}\,K\,T$, so $N_\text{Ar}=\frac{P\,V}{K\,T}$, that is valid as long as $N_\text{Ar}\gg N_\text{X}$. Similarly, we obtain:
\begin{equation}
    \frac{d N_\text{Ar}}{dt} = \frac{P}{K\,T} \, \frac{dV}{dt}.\label{Sol_OR_ctant_1}
\end{equation}
And we can re-express eq. \ref{Sol_OR_ctant_2} as:

\begin{equation}
    f_\text{X}(t) =\frac{Q \cdot A}{P \cdot \Phi} \, \left( 1-\exp \left\{-\frac{\Phi}{V} \cdot t\right\}  \right), \label{Sol_OR_ctant_3}
\end{equation}
where $Q$ is the outgassing rate in units of $[\frac{P \cdot V}{A \cdot t}]$ and $\Phi$ is the flow rate in liters per unit time, as defined in text ($\Phi \equiv \frac{dV}{dt}$). $A$ is the total outgassing area. The formula for the concentration $f_\text{X}(t)$ exhibits a global scaling upon changes of the outgassing rate $Q$, its time dependence being a function of the flow rate and total volume.

\subsection{Power-law outgassing rate}
\label{powerLaw_OR}

A more realistic model for the outgassing of plastics is a power law, where $q(t) \propto t^{-1/2}$. 
The general solution of equation \eqref{eq:mass_flow} is 
\begin{equation}
    N_\text{X}(t) = \exp\left\{-\frac{d N_\text{Ar}}{d t} \cdot  \frac{1}{N_\text{Ar} } \cdot t\right\}
\left(
\int_0^{t} \exp\left\{\frac{d N_\text{Ar}}{d t} \cdot  \frac{1}{N_\text{Ar} } \cdot \tau\right\} \, q(\tau) \, d\tau +c_1\right).
\end{equation}
Assuming again that the concentration is zero at $t=0$ and taking the definition in text, under which $q(t) = {Q_{\text{10h}}} \cdot \frac {A}{KT} \cdot\sqrt{10/t}$, with $t$ in hours, the previous expression can be written as
\begin{equation}
    N_\text{X}(t) = \exp\left\{-\frac{d N_\text{Ar}}{d t} \cdot  \frac{1}{N_\text{Ar} } \cdot t\right\} \, 
\frac{\sqrt{10\pi} \cdot Q_{10\text{h}} \cdot \frac {A}{KT} \cdot\text{erfi}\left( \sqrt{\frac{d N_\text{Ar}}{d t} \frac{1}{N_\text{Ar}}} \sqrt{t} \right)}{\sqrt{\frac{d N_\text{Ar}}{d t} \frac{1}{N_\text{Ar}}}},
\end{equation}
where erfi represents the error function. The concentration of X is then
\begin{equation}
    f_\text{X}(t) = \exp\left\{-\frac{d N_\text{Ar}}{d t} \cdot \frac{1}{N_\text{Ar} } \cdot t\right\} \, 
\frac{\sqrt{10\pi} \cdot Q_{10\text{h}} \cdot \frac {A}{KT} \cdot  \text{erfi}\left( \sqrt{\frac{d N_\text{Ar}}{d t} \frac{1}{N_\text{Ar}}} \sqrt{t} \right)}
{ \sqrt{\frac{d N_\text{Ar}}{d t} \frac{1}{N_\text{Ar}}} \, N_\text{Ar}}.
\end{equation}

Using the definitions of previous section we find:
\begin{equation}
    f_\text{X}(t) = \exp\left\{-\frac{\Phi}{V} \cdot t\right\} \, 
\frac{\sqrt{10\pi} \cdot Q_{10\text{h}} \cdot A \cdot \text{erfi}\left( \sqrt{\frac{\Phi}{V} \cdot t} \right)}
{\sqrt{\frac{\Phi}{V}} \, P \cdot V}.
\end{equation}
($\sqrt{10}$ having units of $[t^{1/2}]$)

\subsection{Gas displacement during filling}
\label{filling}

Under the aforementioned `diffusion-dominated' conditions, it is straightforward to write the evolution equation for contaminant X, assumed to be filling the chamber at $t=0$ (air, for instance), as:

\begin{equation}
    \frac{d N_\text{X}}{d t} = - \frac{d N_\text{Ar}}{d t} \cdot  \frac{N_\text{X}}{N_\text{Ar}}, 
\end{equation} 
yielding the familiar solution:
\begin{equation}
    f_\text{X}(t) = \exp \left\{-\frac{\Phi}{V}\cdot t \right\}.
\end{equation}
        
\subsection{Gas accumulation}
\label{accumulation}

If the outgassing element is oblivious to purification (as assumed in text for N$_2$), eq. \ref{eq:mass_flow} is replaced by:
\begin{equation}
    \frac{d N_\text{X}}{d t} = q(t).
\end{equation} 
The solution is:
\begin{equation}
    f_\text{X}(t) = \frac{K T}{P V} \int_0^t q(t') dt',
\end{equation} 
or
\begin{equation}
    f_\text{X}(t) = \frac{A}{P V} \int_0^t Q(t') dt',
\end{equation} 
as a function of the main magnitudes used in text. When the outgassing decreases with time as $t^{-1/2}$, impurities accumulate in the system as $f_\text{X}\sim t^{1/2}$.

\end{document}